\documentclass[%
 aip,%
 pof,%
 amsmath,amssymb,%
 reprint,%
]{revtex4-2}

\usepackage{graphicx}
\usepackage[caption=false]{subfig}
\usepackage{dcolumn}
\usepackage{bm}

\usepackage{mathrsfs}

\newcommand{\rd}{\mathrm{d}}

\newcommand{\re}{\mathrm{e}}

\newcommand\bnabla{\boldsymbol{\nabla}}
\newcommand\bcdot{\boldsymbol{\cdot}}

\newcommand\Rey{\mbox{Re}}  
\newcommand\Mach{\mbox{Ma}} 
\newcommand\De{\mbox{De}}   
\newcommand\St{\mbox{St}}   
\newcommand\Wo{\mbox{Wo}}   

\usepackage{hyperref}
\hypersetup{
    colorlinks=true,
    breaklinks=true
}

\graphicspath{{figures/}}

\begin{document}

\title[]{Transient compressible flow in a compliant viscoelastic tube}

\author{Vishal Anand}
\thanks{Author to whom correspondence should be addressed.}
\email{anand32@purdue.edu}
\affiliation{School of Mechanical Engineering, Purdue University, West Lafayette, Indiana 47907, USA}

\author{Ivan C.\ Christov}
\email{christov@purdue.edu}
\affiliation{School of Mechanical Engineering, Purdue University, West Lafayette, Indiana 47907, USA}

\date{\today}

\begin{abstract}
Motivated by problems arising in the pneumatic actuation of controllers for micro-electromechanical systems (MEMS), labs-on-a-chip or biomimetic soft robots, and the study of microrheology of both gases and soft solids, we analyze the transient fluid--structure interaction (FSIs) between a viscoelastic tube conveying compressible flow at low Reynolds number. We express the density of the fluid as a linear function of the pressure, and we use the lubrication approximation to further simplify the fluid dynamics problem. On the other hand, the structural mechanics is governed by a modified Donnell shell theory accounting for Kelvin--Voigt-type linearly viscoelastic mechanical response. The fluid and structural mechanics problems are coupled through the tube's radial deformation and the hydrodynamic pressure. For small compressibility numbers and weak coupling, the equations are solved analytically via a perturbation expansion. Three illustrative problems are analyzed. First, we obtain exact (but implicit) solutions for the pressure for steady flow conditions. Second, we solve the transient problem of impulsive pressurization of the tube's inlet. Third, we analyze the transient response to an oscillatory inlet pressure. We show that an oscillatory inlet pressure leads to acoustic streaming in the tube, attributed to the nonlinear pressure gradient induced by the interplay of FSI and compressibility. Furthermore, we demonstrate an enhancement in the volumetric flow rate due to FSI coupling. The hydrodynamic pressure oscillations are shown to exhibit a low-pass frequency response (when averaging over the period of oscillations), while the frequency response of the tube deformation is similar to that of a band-pass filter.
\end{abstract}


\maketitle

\section{Introduction}
\label{sec:intro_transient}

Compressible flow of gases for low \emph{Mach number} $\Mach \ll 1$, yet also low \emph{Reynolds number} $\Rey \ll 1$ have been observed experimentally as early as the 1950s, as explained by \citet{TS57}. Here, the Mach number is defined as $\Mach = \mathcal{V}/c$, where $\mathcal{V}$ is the characteristic flow speed, and $c$ is the speed of sound in the medium. The Reynolds number is defined as $\Rey = \rho \mathcal{V} a/\mu$, where $\rho$ and $\mu$ are the fluid's density and viscosity, respectively, and $a$ is a system length scale.  Nowadays, such low-$\Mach$ and low-$Re$ compressible gas flows find applications at the microscale in micro-electro-mechanical systems (MEMS) \citep{HT96,HT98} for pneumatic flow control \citep{UCTSQ00}, amongst other applications. A hallmark of low-$\Mach$ compressibility in long conduits is the nonlinear axial pressure profile \citep{LTH95,ASB97,ZLLJT02}. As \citet{ASB97} demonstrated, compressible effects in microchannel flows, for $\Rey \ll 1$, can be rationalized by employing lubrication theory in conjunction with a suitable equation of state that accounts for density variations; further theoretical and experimental refinements have followed \citep{LWZ01,ZLLJT02,V06,VB10}. Another early experimental study \citep{WMZTH98} accounted for both the compressibility of the flow and the compliance (bulging) of the flow conduit. However, their theoretical model was not closed (in the sense that it required a fitting parameter to connect the deformed height with the local pressure) nor further validated.

Experimental investigations of compressible flow in channels have proceeded in tandem with the theoretical studies. \citet{PCW86} analyzed the pressure-controlled compressible flow of a monoatomic ideal gas (zero bulk viscosity), to show that compressibility enhances the mass flow rate (compared to the incompressible case) at a fixed pressure drop. \citet{BSG90,BSG93} improved upon the latter results by considering an arbitrary equation of state (accounting for non-zero bulk viscosity), showing that compressibility at low $\Rey$ must be accounted for to ensure the accuracy of capillary viscometers for gases.

A flow is compressible if the density $\rho$ of a fluid particle changes along a pathline. For isothermal flow, density variations are induced solely by pressure variations. The bulk modulus of the fluid  ($\sim \rho c^2$) quantifies the density's resistance to changes caused by pressure forces \citep{KCD16}. If the pressure forces are balanced by the inertial forces in a flow, it is termed a high Reynolds number flow ($\Rey \gg 1)$, and thus the pressure scales as $\rho \mathcal{V}^2$. Then, $\rho \mathcal{V}^2/\rho c^2$, which is the square of $\Mach$, is the dimensionless quantity determining the relevance of compressibility. In engineering, compressibility effects in a flow are considered important\cite{KCD16} when $\Mach > 0.3$.

On the other end of the spectrum, in the low Reynolds number regime, $\Rey \ll 1$, inertial forces ($\sim \rho \mathcal{V}^2$) are negligible compared to viscous forces ($\sim \mu \mathcal{V}/ a$), and $\Mach$ is not a suitable way to quantify compressibility. In the $\Rey\ll1$ regime, the viscous forces balance the pressure forces, meaning that viscous stresses cause significant pressure changes over the length of a flow conduit. Thus, as noted by \citet{S07}, an example of microscale compressible flow emerges in ``the case of gas flows in long channels where the pressure change is sufficiently large that the gas density, which varies in proportion to the pressure, needs to be taken into account'' (Ref.~\onlinecite{S07}, p.~23). In particular, the changes in density are significant for gases because they have low bulk moduli.

Recently, there has been a resurgence of interest in low $\Rey$ compressibility effects, including taking into account microscale fluid--structure interactions (FSIs) \citep{MY18,EJG18,ZFZZC19}. The availability of new polymer-based materials like polydimethylsiloxane (PDMS), a silicon-based polymeric material that allows for rapid and precise manufacture of microdevices \citep{MW02}, coupled with the emergence of new manufacturing methods \citep{C10_book}, such as soft lithography \citep{XW98}, have revolutionized the microfluidics industry \citep{W06}. PDMS is ``soft,'' meaning it has an elastic modulus on the order of a few MPa. Consequently, PDMS-based microchannels and microtubes deform
significantly due to the hydrodynamic pressure forces induced by viscous stresses during flow \cite{KCC18,FZPN19}. The deformation in turn alters the cross-sectional shape, which modifies the velocity profile and, thus, the viscous stress in the flow. In low $\Rey$ turbulent flows in channels with soft walls, this mechanism amplifies the turbulent (Reynolds-averaged) stresses in the flow \cite{RB20}. This feedback loop is an example of the two-way coupled nature of FSIs \citep{F97,P16}. Recent work on incompressible flows in linearly elastic tubes has exploited this feedback loop to create PDMS-based biomimetic \textit{in vitro} models of the microcirculation \cite{KDMCC20}. For incompressible non-Newtonian fluids, on the other hand, it was shown that inlet pressure changes propagate downstream along the elastic tube according to a nonlinear diffusion equation \cite{BBG17}.

FSI in (or around) elastic cylinders (and related \emph{slender} structures) is a time-honored subject in mechanics \citep{P16}. Nevertheless, the current literature on FSIs involving compressible flows has only addressed the case of linearly elastic structures \citep{MY18,EJG18,ZFZZC19}. However, in transient conditions, PDMS is known to exhibit a viscoelastic response \citep{WKB10,RDC19}. In biomechanics, many soft tissues are viscoelastic \citep{F93}. The mechanical properties of lung tissues can be modeled by a linear viscoelastic model in the breathing frequency range \citep{I13}. In particular, the simplest useful model is the Kelvin--Voigt (KV) element consisting of a linearly elastic spring in parallel with a linearly viscous element. Under this modeling approach, \citet{MDLP19} derived  weakly nonlinear governing equations for wave propagation due to incompressible pulsatile flow in a linearly viscoelastic tube. They emphasized that ``viscoelastic effects are very important and should be included in future studies'' (Ref.~\onlinecite{MDLP19}, p.~147), further motivating our work. Specifically, the amplitude of the wave was reduced by $17\%$, when the viscoelasticity of the wall was included, compared to the purely elastic case. However, since their study addressed hemodynamics, compressibility effects were not considered. 

Viscous damping of elastic structures is also a key design feature of soft robotic actuators \citep{SG97a,SG97b,CMLH18}. Damping in a robotic actuator hastens  stabilization by promoting energy dissipation \citep{SG97a}. Specifically, excess energy introduced into the system during the impact of a rigid object on a robotic hand performing grasping must be dissipated to prevent damage to attached sensors. This dissipation of energy can be accomplished by using materials (such as rubbers, gels, sponges etc.) that exhibit viscoelasticity \citep{SG97a,SG97b}. Furthermore, recently, attention has turned to bio-inspired soft robotics \citep{CMLH18}. As noted above, most biological tissues (like skin, muscles, etc.) are viscoelastic \citep{F93}, thus biomimetic soft robots are also designed with a viscoelastic material response in mind \citep{CMLH18}. Most soft robots are actuated pneumatically \citep{KLT13} Thus, understanding \emph{compressible} flows interacting with compliant \emph{viscoelastic} structures  may provide insight into the design of pneumatically actuated biomimetic soft robots. In fact, very recently low $\Rey$ compressible flow FSI has been studied in the context of  actuation of soft robots \citep{BMG20}. So far, however, the FSI literature on viscoelastic \citep{MDLP19} or elastic \citep{SS12,EG14,BBG17,EJG18} tubes with transient (inertial) response has only considered incompressible flow, while the FSI literature on compressible (and variable density) flows has only dealt with linearly elastic structures \citep{ASIDSCB11,EJG18,GSP20}, except for some systems-level modeling of wave speeds for water hammer phenomena in viscoelastic tubes \citep{SW90,WT01} at large $\Rey$. The present work fills a knowledge gap on low $\Rey$ compressible flow in viscoelastic tubes with transient response. 

To this end, we analyze the canonical FSI problem of an initially cylindrical viscoelastic tube conveying the flow of a compressible fluid at low $\Rey$. In Sec.~\ref{sec:problem_formulation}, we introduce the mathematical problem to be solved. In Sec.~\ref{sec:fluid_mechanics}, we derive the governing differential equations which connect the volumetric flow rate with the hydrodynamic pressure and the deformed radius of the tube. In Sec.~\ref{sec:structural_mechanics}, we summarize the governing equations of a Donnell shell theory under the KV model of linear viscoelasticity. Next, the coupled FSI problem is solved via a double perturbation expansion in terms of a FSI parameter and a compressibility number. Three different sub-problems are considered. The first  (Sec.~\ref{sec:steady_state_math}) addresses steady compressible flow in a pressure-drop-controlled linearly elastic tube. The remaining two sub-problems correspond to transient compressible flows. The first transient problem (Sec.~\ref{sec:impulsive}) involves an impulsive pressurization of the tube's inlet, while the second transient problem (Sec.~\ref{sec:oscillatory}) addresses a time-harmonic oscillatory pressure applied at the inlet. Discussion of the results follows in Sec.~\ref{sec:results_discussion}, and conclusions are stated in Sec.~\ref{sec:conclusion}.

The key results of the work are as follows. Both compressibility and FSI enhance the flow rate across the tube. We derive the analytical expression for the dimensionless time constant, which characterizes the transient response of the tube. In the case of oscillatory flow, we show that the tube deformation has a frequency response like that of a band-pass filter, reaching resonance close to the natural frequency of the system. Perhaps the most important finding of this study relates to \emph{acoustic streaming}. We show that, for the case of oscillatory pressure at the inlet, compressibility and FSI work in tandem to generate a streaming flow, even though the main flow is inertialess, similar to peristaltic pumping. The streaming induced enhancement in flow rate, when averaged over the time period of oscillations,  has a frequency response like that of low-pass filter, with a cut-off frequency determined analytically.

\section{Mathematical formulation of the problem} \label{sec:problem_formulation}

\subsection{Preliminaries}
\begin{figure}
\centering
  \includegraphics[width=0.8\linewidth]{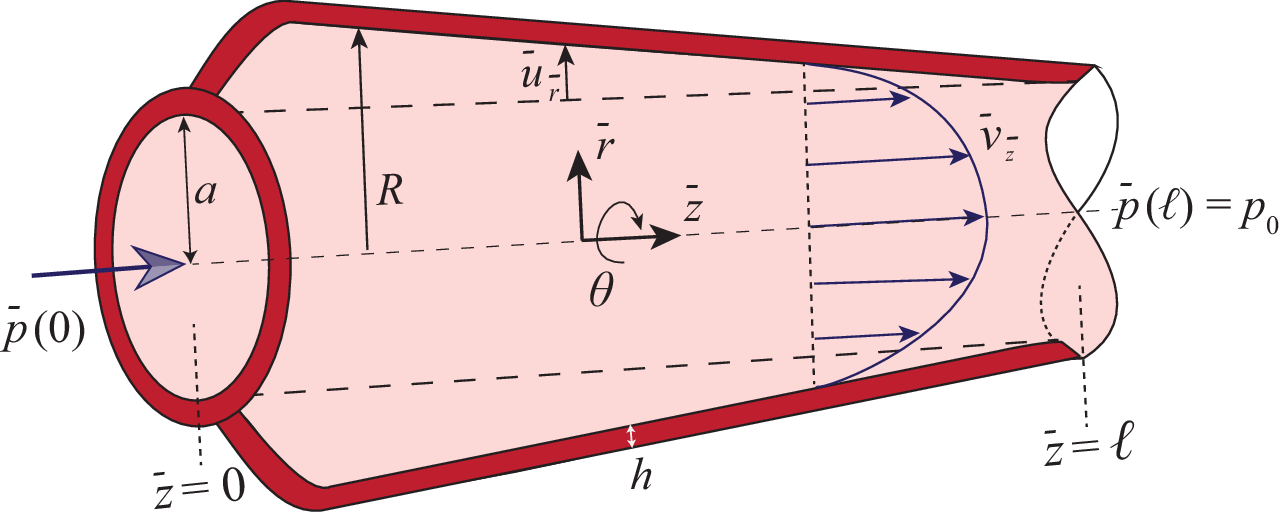}
\caption{Schematic of the microtube geometry. The origin of the coordinate system is at the inlet of the tube, however, it has been displaced from the inlet in the figure for clarity. [Reproduced and adapted with permission from ``On the Deformation of a Hyperelastic Tube Due to Steady Viscous Flow Within,'' Vishal Anand, Ivan C.\ Christov, \textit{Dynamical Processes in Generalized Continua and Structures}, Advanced Structured Materials \textbf{103}, pp.~17--35, doi:10.1007/978-3-030-11665-1\_2.  \textcopyright\ Springer Nature 2019].}
\label{Figure_MC_KV}
\end{figure}

The flow domain of interest consists of an initially cylindrical tube (see Fig.~\ref{Figure_MC_KV}) of undeformed radius $a$, thickness $h$ and length $\ell$. The tube is assumed to be thin $(h \ll a)$ and slender $(a \ll \ell)$. The tube is made of an isotropic, homogeneous, and linearly viscoelastic material that obeys the KV material model, also sometimes referred to as just the `Voigt model' \citep{F93}. The tube is clamped at both its ends ($\bar{z}=0,\ell$).

Both the structural mechanical and fluid mechanical fields are considered axisymmetric in $\theta$. The tube conveys a Newtonian gas. Since this gas flow can be (in general) both unsteady and compressible, the volumetric flow rate $\bar{q}$ is a function of both axial location $\bar{z}$ and time $\bar{t}$, \textit{i.e.}, $\bar{q} = \bar{q}(\bar{z},\bar{t})$. At the inlet, the pressure is imposed as a boundary condition; at the outlet, the pressure is the reference pressure $\bar{p}=p_0$. Due to the hydrodynamic pressure exerted by the transient compressible flow, the tube deforms, and the deformed radius is  $\bar{R}(\bar{z},\bar{t}) =a +\bar{u}_{\bar{r}}(\bar{z},\bar{t})$, where $\bar{u}_{\bar{r}}$ is the radial displacement of the structure. 

Through a perturbative approach,  we aim to establish a predictive relationship between the key fluid mechanical and structural mechanical quantities: $\bar{q}(\bar{z},\bar{t})$, $\bar{p}(\bar{z},\bar{t})$, $\bar{R}(\bar{z},\bar{t})$, and $\bar{u}_{
\bar{r}}(\bar{z},\bar{t})$, as well as their spatiotemporal variations due to unsteadiness.
 
\subsection{Fluid mechanics}
\label{sec:fluid_mechanics}

The fluid mechanics problem will analyzed under the following assumptions:
\begin{enumerate}
   \item Newtonian gas with zero bulk viscosity (exact for monoatomic gases).\footnote{The assumption of zero bulk viscosity is not crucial as the terms in the momentum equation affected by bulk viscosity are negligible under the lubrication approximation \citep{ASB97,EJG18}.}
   \item Axisymmetric flow without swirl: ${\partial}(\,\cdot\,)/{\partial \theta} = 0$ and $v_{\theta} = 0$.
   \item Slender tube: $\ell \gg a \;\Leftrightarrow\; \epsilon = {a}/{\ell} \ll 1$.
   \item Isothermal flow.
   \item A linear equation of state connects the density and pressure. 
\end{enumerate}

\subsubsection{Continuity equation}
\label{sec:fluid_cont_eq}

We introduce the following dimensionless variables: 
\begin{equation}
    t = \bar{t}/\mathcal{T}_f, \quad
    r = \bar{r}/a, \quad
    z = \bar{z}/\ell, \quad
    v_{z} = \bar{v}_{\bar{z}}/\mathcal{V}_z, \quad
    v_{r} = \bar{v}_{\bar{r}}/\mathcal{V}_r,\quad
    \rho = \bar{\rho}/\rho_0, \quad
    p = \left(\bar{p}-p_0\right)/\mathcal{P}_c,
\label{eq:nd_vars_tube}
\end{equation}
where $\bar{p}$ is the absolute pressure. Here, $\mathcal{P}_c$ is the characteristic pressure scale to be determined  from the boundary condition, since we study a pressure-controlled system. $\mathcal{V}_z$ and $\mathcal{V}_r$ are characteristic scales for the axial and radial velocities, respectively, connected by the conservation of mass, and related to $\mathcal{P}_c$ by the conservation of momentum equation. Finally, $\mathcal{T}_f$ is characteristic time scale, and $\rho_0$ is a characteristic scale for the density, which is the density at the outlet, where the (absolute) reference pressure is $p_0$.

For axisymmetric, transient flow of a gas, the equation of continuity in cylindrical coordinates takes the form \citep{panton,KCD16}:
\begin{equation}
    \frac{\partial \bar{\rho}}{\partial \bar{t}}+\frac{1}{\bar{r}}\frac{\partial}{\partial \bar{r}}(\bar{r} \bar{\rho} \bar{v}_{\bar{r}}) + \frac{\partial}{\partial \bar{z}}(\bar{\rho} \bar{v}_{\bar{z}}) = 0,
    \label{eq:compressible_com}
\end{equation}
where $\bar{\rho}(\bar{r},\bar{z},\bar{t})$ is the density of the gas and $\bm{v}(\bar{r},\bar{z},\bar{t}) = (\bar{v}_{\bar{r}},0,\bar{v}_{
\bar{z}})$ is its axisymmetric velocity field.  Substituting the variables from Eqs.~\eqref{eq:nd_vars_tube} into Eq.~\eqref{eq:compressible_com} yields
\begin{equation}
\label{eq:continuity_eqn_dimless}
   \frac{\rho_0}{\mathcal{T}_f} \frac{\partial \rho}{\partial  t} + \frac{\rho_0 \mathcal{V}_r}{a} \frac{1}{r} \frac{\partial}{\partial   r}(\rho   r   v_{r}) + \frac{\rho_0 \mathcal{V}_z}{\ell}\frac{\partial }{\partial z} (\rho v_{z}) = 0.
\end{equation}
Balancing the latter equation (in an order-of-magnitude sense), we obtain 
\begin{equation}
    \mathcal{V}_r = \epsilon \mathcal{V}_z,\qquad \mathcal{T}_f = \ell/\mathcal{V}_z.
\label{eq:scales_related}
\end{equation}

\subsubsection{Momentum equations}
\label{sec:fluid_mom_eq}

For the compressible flow of a Newtonian gas with zero bulk viscosity, the conservation of linear momentum equation in the axial $z$-direction takes the form \citep{panton,KCD16}:
\begin{equation}
    \bar{\rho}\left(\frac{\partial \bar{v}_{\bar{z}}}{\partial \bar{t}}+\bar{v}_{\bar{r}}\frac{\partial \bar{v}_{\bar{z}}}{\partial \bar{r}}+\bar{v}_{\bar{z}}\frac{\partial \bar{v}_{\bar{z}}}{\partial \bar{z}}\right)
    = \frac{1}{\bar{r}}\frac{\partial}{\partial \bar{r}}\left(\mu \bar{r}\frac{\partial \bar{v}_{\bar{z}}}{\partial \bar{r}}+\mu \bar{r} \frac{\partial \bar{v}_{\bar{r}}}{\partial \bar{z}}\right)    +\frac{\partial }{\partial \bar{z}}\left(2\mu\frac{\partial \bar{v}_{\bar{z}}}{\partial \bar{z}}-\frac{2}{3}\mu\bar{\bnabla}\bcdot\bar{\bm{v}}\right)-\frac{\partial \bar{p}}{\partial \bar{z}}.
    \label{eq:z_momentum}
\end{equation}
By using the scales introduced in Eqs.~\eqref{eq:nd_vars_tube} and \eqref{eq:scales_related}, the dimensionless form of Eq.~\eqref{eq:z_momentum} is 
\begin{multline}
\label{eq:z_momentum_dimless}
    \frac{\rho_0 \mathcal{V}_z^2}{\ell}\rho\left(\frac{\partial  v_{z}}{\partial  t} + v_{r}\frac{\partial   v_{z}}{\partial  r} +  v_{z}\frac{\partial   v_{z}}{\partial z}\right)\\
    =\frac{\mu \mathcal{V}_{z}}{a^2}\frac{1}{r}\frac{\partial}{\partial   r}\left(  r\frac{\partial   v_{z}}{\partial   r} + \epsilon^2  r\frac{\partial   v_{r}}{\partial z}\right)  + \frac{\mu \mathcal{V}_{z}}{\ell^2}\frac{\partial }{\partial z}\left(2\frac{\partial   v_{z}}{\partial z}-\frac{2}{3}{\bnabla}\bcdot{\bm{v}}\right) - \frac{\mathcal{P}_c}{\ell}\frac{\partial p}{\partial z}.
\end{multline}

Since the flow is compressible (and isothermal), a constitutive equation must be specified to relate $\bar{\rho}$ and $\bar{p}$. We use a linear equation of state \citep{V06, VB10}:
\begin{equation}
    \bar{\rho} =\rho_{0}\left(1+\frac{\bar{p}-p_0}{\mathcal{B}_T}\right),\qquad \mathcal{B}_T = \rho_0\left(\frac{\partial \bar{p}}{\partial \bar{\rho}}\right)_T,
\label{eq:constitutive_eq}
\end{equation}
where $\mathcal{B}_T = const.$ is the isothermal bulk modulus of the gas. The dimensionless form of Eq.~\eqref{eq:constitutive_eq} is then
\begin{equation}
    \rho = 1+ \alpha p,\qquad \alpha = \mathcal{P}_c/\mathcal{B}_T,
\label{eq:constitutive_eq_dimless}
\end{equation}
where the dimensionless parameter $\alpha$ is termed the \emph{compressibility number}. The chosen equation of state~\eqref{eq:constitutive_eq_dimless} (and its dimensional counterpart, Eq.~\eqref{eq:constitutive_eq}) can be interpreted as a Taylor-series expansion of the density about the outlet value, in terms of the gauge pressure. Compressibility is important in the flow if $\mathcal{P}_c\sim\mathcal{B}_T$, \textit{\textit{i.e.}}, $\alpha = \mathcal{O}(1)$. Therefore, $\alpha$ fulfills the same role, for low-$\Rey$ viscous flow, as $\Mach$ does for high-$\Rey$ inertial flow.\footnote{Indeed, we could also express $\alpha$ in terms of the adiabatic speed of sound $c=c_0=\sqrt{\gamma \mathcal{B}_T/\rho_0}$ as $\alpha = \gamma \mathcal{P}_c/\rho_0 c_0^2$, where $\gamma$ is the ratio of specific heats.}

Using Eq.~\eqref{eq:constitutive_eq_dimless},  Eq.~\eqref{eq:z_momentum_dimless} can be rewritten as:
\begin{multline}
\label{eq:z_momentum_dimless2}
    \epsilon \Rey(1+\alpha p)\left(\frac{\partial   v_{z}}{\partial  t}+  v_{  r}\frac{\partial   v_{z}}{\partial   r}+  v_{z}\frac{\partial   v_{z}}{\partial z}\right)\\
    =\frac{1}{r}\frac{\partial}{\partial   r}\left(r\frac{\partial  v_{z}}{\partial  r} + \epsilon^2 r\frac{\partial  v_{  r}}{\partial z}\right) +\epsilon^2\frac{\partial }{\partial z}\left(2\frac{\partial   v_{z}}{\partial z}-\frac{2}{3}{\bnabla}\bcdot{\bm{v}}\right) - \frac{\mathcal{P}_c a^2}{\ell \mathcal{V}_z \mu}\frac{\partial p}{\partial z},
\end{multline}
where $\Rey = {\rho_0\mathcal{V}_z a}/{\mu}$ is the Reynolds number. Dropping terms of $\mathcal{O}(\epsilon^2)$ and $\mathcal{O}({\epsilon} \Rey$),  Eq.~\eqref{eq:z_momentum_dimless2} reduces to:
\begin{equation}
\label{eq:z_momentum_dimless3}
  0  =\frac{1}{  r}\frac{\partial}{\partial   r}\left(  r\frac{\partial   v_{z}}{\partial   r}\right) -\frac{\mathcal{P}_ca^2}{\ell \mathcal{V}_z \mu} \frac{\partial p}{\partial z}.
\end{equation}
Balancing the last equation yields the axial velocity scale:
\begin{equation}
\mathcal{V}_z = \frac{\mathcal{P}_c a^2}{\mu \ell}.
\label{eq:velocity_scale}
\end{equation}
We study pressure-controlled systems in which $\mathcal{P}_c$ is set by the boundary conditions, so $\mathcal{V}_z$ can be, in principle,  calculated from Eq.~\eqref{eq:velocity_scale}.\footnote{We can also consider a system in which the inlet mass flow rate $\dot{m}$ is specified. Then, $\mathcal{V}_z \equiv \dot{m}/ (\rho_{0}\pi a^2)$, in which case $\mathcal{P}_c$ can be, in principle, calculated from Eq.~\eqref{eq:velocity_scale}.}
Thus, Eq.~\eqref{eq:z_momentum_dimless3} is finally written as:
\begin{equation}
\label{eq:z_momentum_dimless4}
  0  =\frac{1}{  r}\frac{\partial}{\partial   r}\left(  r\frac{\partial   v_{z}}{\partial   r}\right) -\frac{\partial p}{\partial z}.
\end{equation}

Using  Eqs.~\eqref{eq:nd_vars_tube}, \eqref{eq:scales_related}, \eqref{eq:constitutive_eq_dimless}, and \eqref{eq:velocity_scale}, the dimensionless $r$-momentum equation \citep{panton,KCD16} is
\begin{multline}
    \epsilon^3 \Rey(1+\alpha p)\left(\frac{\partial v_{r}}{\partial  t}+  v_{r}\frac{\partial v_{r}}{\partial r} +  v_{z} \frac{\partial v_{  r}}{\partial z}\right)\\
    = \epsilon^2 \frac{1}{r}\frac{\partial}{\partial   r}\left(2 r\frac{\partial v_{r}}{\partial r} - \frac{2}{3}   r{\bnabla}\bcdot{\bm{v}}\right)+\epsilon^2\frac{\partial }{\partial z}\left(\frac{\partial  v_{z}}{\partial  r}+\epsilon^2\frac{\partial  v_{ r}}{\partial z}\right)-\frac{\partial p}{\partial  r}.
    \label{eq:r_momentum_dimless}
\end{multline}
Neglecting small terms, we obtain:
\begin{equation}
\label{eq:r_momentum_dimless_3}
    0 = \frac{\partial p}{\partial r}.
\end{equation}

\subsubsection{Velocity boundary conditions}
We impose the traditional no-slip boundary condition \citep{LBS07} on the axial and radial velocities at the deformed tube wall:
\begin{align}
       v_z|_{r=R} &= 0,
  \label{eq:velocity_bc_axial} \\
       v_{r}|_{r=R} &= \left.\beta\left(\frac{\partial u_{r}}{\partial  t}+  v_{z}\frac{\partial u_{r}}{\partial z}\right)\right|_{r=R}.
     \label{eq:velocity_bc_radial}  
\end{align}
Observe that, since the tube wall deformation is transient, Eq.~\eqref{eq:velocity_bc_radial} is, in fact, the kinematic boundary condition \citep{panton}, where ${u}_{r}$ is the (dimensionless) radial deformation of the tube, and we have neglected axial displacements (to be justified in Sec.~\ref{sec:structural_mechanics} below). 
Here, $\beta = \mathcal{U}_c/a$ can be termed the \emph{FSI coupling parameter}; $\mathcal{U}_c$ is the characteristic deformation scale, which is to be determined upon analyzing the mechanical force balance on the structure (in Sec.~\ref{subsection:DKV_model}, just before Eq.~\eqref{eq:Deformation_ODE_final}). Then, the deformed radius $\bar{R}$, can be written in terms of the dimensionless variables as:
\begin{equation}
    R(z,t) = \frac{\bar{R}(\bar{z},\bar{t})}{a} = \frac{a+{\bar{u}}_{\bar{r}}(\bar{z},\bar{t})}{a} = 1+\beta u_{r}(z,t).
\end{equation}

\subsubsection{Velocity profile}
\label{sec:vel_field}
We solve Eq.~\eqref{eq:z_momentum_dimless3} subject to Eq.~\eqref{eq:velocity_bc_axial} to obtain the axial velocity field:
 \begin{equation}
       v_{z}(r,z,t) = -\frac{1}{2}\frac{\partial p}{\partial z} \left[\frac{(1+\beta u_{r})^2 - r^2}{2}\right].
     \label{eq:velocity_axial_dimless}
 \end{equation}
Observe the flow is primarily axial but it is  two-dimensional. Here, $\beta$, $u_{r}(z,t)$, and $p(z,t)$ are independent of $r$.
The volumetric flow rate is the area integral of the axial velocity from Eq.~\eqref{eq:velocity_axial_dimless}:
\begin{equation}
\label{eq:flow_rate_defined}
q(z,t) \equiv \frac{\bar{q}(\bar{z},\bar{t})}{\mathcal{V}_z\pi a^2}  =  \int\limits_{0}^{R(z, t)}  v_z(r,z,t) \, 2r \,\rd r =  -\frac{\partial p}{\partial z}\left[\frac{1}{8}(1+\beta  u_{  r})^4\right].
\end{equation}
As a consistency check, observe that for $\beta\to0$, Eq.~\eqref{eq:flow_rate_defined} reduces to $ q= -(1/8){\partial p}/{\partial z}$, which is the Hagen--Poiseuille law in dimensionless form \citep{SS93}. 

In previous studies of steady incompressible flow and FSI in tubes \citep{AC18b}, the volumetric flow rate $q$ was specified. Then, by conservation of mass $q=const$ throughout the tube and, therefore,  Eq.~\eqref{eq:flow_rate_defined} is simply an ordinary differential equation (ODE) in $p(z)$.  Here, however, we deal with unsteady compressible flow, thus $q$ from Eq.~\eqref{eq:flow_rate_defined} is \emph{not} constant, \textit{i.e.}, $q = q(z,t)$. 

\subsubsection{Unsteady volumetric flow rate and time scales}

To determine the governing equation for $q$, we integrate the continuity Eq.~\eqref{eq:continuity_eqn_dimless} across the radial extent of the tube. Upon using the boundary conditions from Eqs.~\eqref{eq:velocity_bc_axial} and \eqref{eq:velocity_bc_radial}, we obtain
\begin{equation}
\label{eq:continuity_equation_integrated}
   \frac{\partial}{\partial t} \left[ \rho(1+\beta  u_{r})^2 \right] + \frac{\partial (\rho q)}{\partial z}=0.
\end{equation}
Observe that the unsteady term in Eq.~\eqref{eq:continuity_equation_integrated} can be written out as
\begin{multline}
 \frac{\partial}{\partial t} \left[ \rho(1+\beta  u_{r})^2 \right] \\
 = \underbrace{\left(1+\frac{\bar{u}_{\bar{r}}}{a}\right)^2}_{\mathcal{O}(1)} \frac{\partial }{\partial \bar{t}}\underbrace{\left(\frac{\bar{p}-p_0}{\mathcal{P}_c}\right)}_{\mathcal{O}(1)}\left(\alpha \mathcal{T}_f\right) + 2\underbrace{\left(1+\frac{\bar{p}-p_0}{\mathcal{B}_T}\right)}_{\mathcal{O}(1)}\underbrace{\left(1+\frac{\bar{u}_{\bar{r}}}{a}\right)}_{\mathcal{O}(1)}\frac{\partial }{\partial \bar{t}}\underbrace{\left(\frac{\bar{u}_{\bar{r}}}{\mathcal{U}_c}\right)}_{\mathcal{O}(1)}\left(\beta \mathcal{T}_f\right).
 \label{eq:Terms}
\end{multline}
From the first term on the right-hand side of Eq.~\eqref{eq:Terms}, we deduce that to retain the transient response of the pressure changes in the balanced continuity equation~\eqref{eq:continuity_equation_integrated}, the time scale of compressibility must be $\mathcal{T}_\mathrm{compressibility} \sim \alpha \mathcal{T}_f = \alpha\ell/\mathcal{V}_z $. Likewise, from the second term on the right-hand side of Eq.~\eqref{eq:Terms}, we deduce that  $\mathcal{T}_\mathrm{FSI} \sim \beta \mathcal{T}_f = \beta\ell/\mathcal{V}_z$ is the time scale of FSI (transient deformation response). The latter is same time scale deduced in prior work on microchannels \citep{MCSPS19} and microtubes  \citep{EG14}. 

Thus, we have shown that transient terms arise in the continuity equation due acceleration of the fluid, FSI and compressibility. From Eq.~\eqref{eq:z_momentum_dimless2}, the acceleration of the fluid $\sim \epsilon Re$, while from Eq.~\eqref{eq:Terms} the compressibility effects $\sim\alpha$, and the FSI effects $\sim \beta$. The fact that we have chosen to neglect the acceleration of the fluid, but kept the transient compressibility and FSI, leads us to the following requirement on the dimensionless parameters:
\begin{equation}
    \epsilon \Rey \ll \beta\ll1,\qquad \epsilon \Rey \ll \alpha \ll 1.
\label{eq:scales_of_flow}
\end{equation}
These conditions must be satisfied under our quasi-steady lubrication theory. Specifically, since both $\alpha$ and $\beta$ are small, we are able to use a double perturbation expansion in $\alpha$ and $\beta$ to solve the governing equations in later sections. Moreover, observe that Eq.~\eqref{eq:scales_of_flow} does not yield an ordering between the small parameters $\alpha$ and $\beta$. Thus, given the lack of \emph{a priori} scale separation, all quadratic terms in the perturbation expansions that follow are neglected.

Equations~\eqref{eq:flow_rate_defined} and \eqref{eq:continuity_equation_integrated} contain three unknowns, namely $u_{r}$, $p$, $q$. We need another equation to uniquely determine these quantities. To that end, we turn our attention to the structural mechanics problem.

\subsection{Structural mechanics}
\label{sec:structural_mechanics}

The key assumptions pertaining to the structural mechanical aspect of the FSI problem are:
\begin{enumerate}
    \item The tube is clamped at both ends.
    \item The tube is slender: its undeformed radius is small compared to its length ($a\ll \ell$). The tube is thin: its thickness is small compared to its undeformed radius ($h\ll a$).
    \item The tube is assumed to be in a state of plane strain, effectively decoupling each axial cross-section from the next \cite{AC18b}, owing to the previous two assumptions.
    \item The tube is composed of a linearly viscoelastic material obeying the Kelvin--Voigt model of viscoelasticity.
    \item The deformations and strains are small.
    \item The load on the structure and the deformation field are both axisymmetric.
\end{enumerate}

To formulate the equations governing the deformation of the tube, we use Donnell shell theory \cite{DonnellShell,Kraus67,D90}. However, since the (classical) Donnell shell theory is valid only for linearly elastic material, it will be modified to account for Kelvin--Voigt viscoelasticity.

\subsubsection{Linearly elastic Donnell shell}

Before we start our exposition of Donnell shell theory, it is pertinent to mention that, owing to assumptions 1--3 above, the (normal) axial strain can be negligible, or
\begin{equation}
    \frac{\partial \bar{u}_{\bar{z}}}{\partial \bar{z}} = 0 \qquad\Rightarrow\qquad \bar{u}_{\bar{z}} \equiv 0.
\end{equation}
Next, according to the Donnell shell theory, the normal stress resultant in the circumferential direction ($= \int_a^{a+h}\bar{\sigma}_{\bar{\theta}\bar{\theta}}\,\rd \bar{r}$) are\citep{D90,Kraus67}
\begin{subequations}\begin{align}
\label{eq:NormalStressResultant}
    \bar{N}_{\theta\theta} &= h \hat{E}\left( \frac{\bar{u}_{\bar{r}}}{a}+\frac{1}{a}\frac{\partial \bar{u}_{\bar{\theta}}}{\partial 
    \bar{\theta}}+\nu\frac{\partial \bar{u}_{\bar{z}}}{\partial \bar{z}}\right),\\
     & = h \hat{E}\left( \frac{\bar{u}_{\bar{r}}}{a}\right),
\end{align}\label{eq:Donnell_const_eq_1}\end{subequations}
where the second equation follows from the assumptions of axisymmetry and plane strain.
     
Similarly, the expression for bending moment ($= \int_a^{a+h}\bar{\sigma}_{\bar{z}\bar{z}} \bar{r} \,\rd \bar{r}$) is
\begin{subequations}
\begin{align} 
\label{eq:BendingMoment}
   \bar{M}_{\bar{z}\bar{z}} &= -\frac{h^3}{12}\hat{E}\left(\frac{\partial^2\bar{u}_{\bar{r}}}{\partial \bar{z}^2}+\frac{\nu}{a^2}\frac{\partial ^2 \bar{u}_{\bar{r}}}{\partial \bar{\theta^2}}\right), \\
   &=-\frac{h^3}{12}\hat{E}\left(\frac{\partial^2\bar{u}_{\bar{r}}}{\partial \bar{z}^2}\right),
\end{align}\label{eq:Donnell_const_eq_2}\end{subequations}
where the second equation follows from the assumption of axisymmetry. Here, $\hat{E} = E/(1-\nu^2)$ is the plane strain Young's modulus, and $\nu$ is the Poisson ratio.

Meanwhile, the equation of equilibrium in the radial direction \citep{D90,Kraus67} is
\begin{equation}
    \frac{\partial^2 \bar{M}_{\bar{z}\bar{z}}}{\partial \bar{z}^2}-\frac{\bar{N}_{\theta\theta}}{a} +\bar{p} = {\rho}_s h\frac{\partial ^2 \bar{u}_{\bar{r}}}{\partial \bar{t}^2},
    \label{eq:Equilibrium}
\end{equation}
where $\bar{p}$ is the radial load on the structure due to the fluid flow within, keeping in mind that $\bar{p}=\bar{p}(\bar{z},\bar{t})$, and ${\rho}_s$ is the constant density of the solid material.

\subsubsection{Linearly viscoelastic Donnell shell}
\label{subsection:DKV_model}

The correspondence between the constitutive equations of linear elasticity and the KV model of viscoelasticity \citep{F75} allows us to analogously write the versions of the constitutive relations~\eqref{eq:Donnell_const_eq_1} and \eqref{eq:Donnell_const_eq_2}, for a viscoelastic Donnell shell, as\citep{BSW95,BDS00}:
\begin{subequations}\label{eq:NM_Lame_Visco_2}\begin{align}
\label{eq:NormalStressResultant_Visco}
  \bar{N}_{\theta\theta} &= \left(\frac{h}{a}\right) \hat{E}{\bar{u}_{{\bar{r}}}}+\left(\frac{h}{a}\right) C_v\frac{\partial \bar{u}_{\bar{r}}}{\partial \bar{t}},\\
\label{eq:BendingMoment_Visco}
   \bar{M}_{\bar{z}\bar{z}} &= -\frac{h^3}{12}\hat{E}\frac{\partial^2 \bar{u}_{\bar{r}}}{\partial \bar{z}^2} - \frac{h^3}{12}C_v \frac{\partial^3\bar{u}_{\bar{r}}}{\partial \bar{t} \partial \bar{z}^2}.
\end{align}\label{eq:visco_Donnell_const_eq}\end{subequations}
In short, the corresponding linearly viscoelastic Donnell shell's constitutive equations are obtained by applying the relaxation operator $1 + (C_v/\hat{E})\partial/\partial \bar{t}$ to the constitutive equations~\eqref{eq:Donnell_const_eq_1} and ~\eqref{eq:Donnell_const_eq_2}. 
Here, $C_v$ is a viscoelastic modulus analogous to  $\hat{E}$. 

Substituting Eqs.~\eqref{eq:NM_Lame_Visco_2}  into the equilibrium equation~\eqref{eq:Equilibrium}, yields the following transient partial differential equation (PDE) for the radial deformation of a thin cylindrical tube made from a linearly viscoelastic material\citep{PAG17}:
\begin{equation}
\label{eq:ViscoElastic_Donnell}
  \frac{h^3}{12}\hat{E}{\frac{\partial^4\bar{u}_{\bar{r}}}{\partial \bar{z}^4}}+\frac{h^3}{12}C_v{\frac{\partial^5\bar{u}_{\bar{r}}}{\partial {\bar{t}} \partial \bar{z}^4}}+ {\frac{h}{a^2}}{\hat{E}}\bar{u}_{\bar{r}}+\frac{h}{a^2} C_v\frac{\partial \bar{u}_{\bar{r}}}{\partial {\bar{t}}}+\rho_s h\frac{\partial ^2 \bar{u}_{\bar{r}}}{\partial {\bar{t}}^2}=\bar{p}.
\end{equation}
Next, we make the governing equation~\eqref{eq:ViscoElastic_Donnell} dimensionless by introducing the dimensionless variables from Eq.~\eqref{eq:nd_vars_tube}:
\begin{equation}
    \label{eq:Deformation_Profile_1}
    \underbrace{\frac{\mathcal{E}^4}{12} \frac{\partial^4 u_{r}}{d z^4}}_{\text{bending}} + \underbrace{\frac{\mathcal{E}^4}{12\De} \frac{\partial^5 u_{r}}{\partial z^4 \partial  t}}_{\text{damping}} + \underbrace{u_{r}}_{\text{stretching}} + \underbrace{\frac{1}{\De} \frac{\partial u_{r}}{\partial  t}}_{\text{damping}}
    +\underbrace{\St\, \frac{\partial^2  u_{r}}{\partial t^2}}_{\text{inertia}}  = \underbrace{\frac{\mathcal{P}_c a^2 }{\hat{E}  h \mathcal{U}_c}p}_{\text{loading}},
\end{equation}
where $\mathcal{U}_c$ is the characteristic deformation scale to be determined.

There are three dimensionless numbers that determine the evolution of the deformation: $\mathcal{E}$, $\De$, and $\St$. First, $\mathcal{E} = \sqrt{{ha}/{\ell^2}}$ is a dimensionless number that determines the width of the bending boundary layers in the tube near its clamped edges at $z=0,1$. For example, $\mathcal{E}\ll1$ means that most of the tube is in a stretching state, while $\mathcal{E}\gg1$ corresponds to the case of a tube in a mostly bending-dominated state \citep{D90}.  Second, the Deborah number\citep{R64} $\De = {\hat{E}\mathcal{T}_f}/{C_v}$  quantifies the viscoelastic response of the tube. For example, for $\De\gg1$, the tube behaves like an elastic solid, while  for $\De\ll1$  the viscous/damping response of the tube dominates. Third, the Strouhal number $\St= {\rho_s a^2}/(\mathcal{T}_f^2 \hat{E})$ quantifies the inertial response of the tube; for $\St\ll1$, the inertial response of the tube is negligible.

The unknown scale of elastic deformation $\mathcal{U}_c $ is now chosen to make the coefficient of the right-hand side of Eq.~\eqref{eq:Deformation_Profile_1} equal to unity: $\mathcal{U}_c = \mathcal{P}_ca^2/(\hat{E}h)$, which is as in previous work on the steady problem \citep{AC18b}, except here we have incorporated the factor $(1-\nu^2)$ into $\hat{E}$ for convenience. Next, we neglect bending (\textit{i.e.}, $\mathcal{E}^4 \ll 1\Rightarrow \mathcal{E}^4/ \De\ll 1$), which is justified by the prior assumptions of slenderness and thinness \citep{D90,CGM07}. Then, the final form of the governing PDE for the radial deformation of the thin viscoelastic tube under hydrodynamic loading is:
\begin{equation}
\label{eq:Deformation_ODE_final}
u_{r}+ \frac{1}{\De}\frac{\partial u_{r}}{\partial t} + \St\frac{\partial ^2  u_{r}}{\partial  t^2} = p.
\end{equation}
with the following initial conditions:
\begin{equation}
    \label{eq:Initial_Condition}
    u_r(z, t= 0) = 0 \qquad \frac{\partial u_r}{\partial t}(z, t = 0) = 0.
\end{equation}

Importantly, observe that the Eq.~\eqref{eq:Deformation_ODE_final} has a proper linearly elastic balance in the limit $\De\to\infty$. Similarly to the analysis of Eq.~\eqref{eq:Terms}, we also deduce from Eq.~\eqref{eq:Deformation_ODE_final} that the tube's damping time scale is $\mathcal{T}_{\text{damping}} \sim \mathcal{T}_f/\De$, while the tube's inertial time scale is $\mathcal{T}_{\text{inertial}} \sim \sqrt{\St}\mathcal{T}_f$.

\subsection{Summary of the model's governing equations}

The fluid mechanics problem is governed by Eqs.~\eqref{eq:flow_rate_defined} and \eqref{eq:continuity_equation_integrated}, and the structural mechanics problem is governed by Eq.~\eqref{eq:Deformation_ODE_final}. These are supplemented by the equation of state~\eqref{eq:constitutive_eq_dimless}. This set of four coupled partial differential and algebraic equations governs the evolution of fluid's density $\rho(z,t)$, volumetric flow rate $q(z,t)$, hydrodynamic pressure $p(z,t)$, and the radial deformation  $u_r(z,t)$ of the tube. The model's dimensionless parameters are summarized in Table~\ref{tbl:Dimless_Parameters}, along with their relative orders of magnitude, wherever applicable.

The next step in our analysis is to solve the governing equations and analyze the dynamics of the transient FSI problem of compressible flow in a viscoelastic tube. 

\begin{table}
\centering
  \begin{tabular}{@{\extracolsep{1em}}llll}
    \hline\hline
    Parameter & Name & Definition & Magnitude\\
    \hline
    $\epsilon$ & Aspect ratio & $a/\ell$ & $\ll 1$ \\
    $\epsilon \Rey$ & Reduced Reynolds number & $\rho_0 \mathcal{V}_z a^2/(\mu\ell)$ & $\ll 1$\\
    $\alpha$ & Compressibility number & $\mathcal{P}_c/\mathcal{B}_T$ & $\epsilon \Rey  \ll \alpha \ll 1$\\
    $\beta$ & FSI coupling parameter & $\mathcal{P}_c a/(\hat{E}h)$ & $\epsilon \Rey \ll \beta \ll 1$\\
    $\mathcal{E}$ & Bending parameter & $\sqrt{ha/\ell^2}$ & $\mathcal{E}^4 \ll 1$\\
    $\De$ & Deborah number & $\hat{E}\mathcal{T}_f/C_v$ & $> 1/(2\sqrt{\St})$\\
    $\St$ & Strouhal number & $\rho_s a^2/(\mathcal{T}_f^2 \hat{E})$ & $=\mathcal{O}(1)$\\
    \hline\hline
\end{tabular}
\caption{Dimensionless parameters that govern the FSI problem, their mathematical definitions, and their relative orders of magnitude.}
\label{tbl:Dimless_Parameters}
\end{table}

\section{Exact and perturbative solutions to the coupled problem}

\subsection{Steady response}
\label{sec:steady_state_math}

First, consider the steady-state problem, wherein the linearly viscoelastic tube model reduces to a linearly elastic tube model. Setting $\partial(\,\cdot\,)/\partial t=0$ in  Eq.~\eqref{eq:continuity_equation_integrated}, we find that the mass flow rate in any tube cross-section is the same constant:
\begin{equation}
 \frac{\partial (\rho q)}{\partial z}=0 \quad\Rightarrow\quad \rho q = \dot{m}_0=const.
\label{eq:ss_governing_equation}
\end{equation}
Note that, unlike the case of an imposed inlet mass flow rate, here $\dot{m}_0$ is an \emph{unknown} constant, to be determined from the boundary conditions, as a function of the imposed pressure drop.

The structural mechanics equation~\eqref{eq:Deformation_ODE_final} at steady state is simply a linear deformation--pressure relation: $u_{r}(z) =  p(z)$. Substituting this relation, $q$ from Eq.~\eqref{eq:flow_rate_defined} and $\rho$ from Eq.~\eqref{eq:constitutive_eq_dimless} into Eq.~\eqref{eq:ss_governing_equation} and using the inlet boundary condition $\bar{p}(0) = \mathcal{P}_c$, \textit{i.e.}, $p(0) = 1$, leads to an ordinary differential equation (ODE) in $p(z)$:
\begin{equation}
\label{eq:pressure_steady_general}
    (1+\beta)^5\left [(6+5\alpha)\beta-\alpha\right] -[1+\beta p(z)]^5\left \{[6+ 5\alpha p(z)]\beta -\alpha \right\} = 240 \beta^2 \dot{m}_0 z .
\end{equation}
This equation describes the pressure variation due to FSI between a compressible flow and a linearly elastic tube. In this case,  $\dot{m}_0$ is found from Eq.~\eqref{eq:pressure_steady_general} by imposing the usual boundary condition at the outlet: gauge pressure, $p(1) = 0$. For the case of $\alpha=0$, Eq.~\eqref{eq:pressure_steady_general} reduces to the pressure-controlled equivalent of the flow-rate-controlled expression for an incompressible flow obtained by \citet{AC18b}. 

\subsection{Impulsively pressurization of the inlet}
\label{sec:impulsive}

Next, we solve the transient problem in which the flow is driven by an impulsive pressure at the inlet boundary condition. To this end, first Eqs.~\eqref{eq:flow_rate_defined} and \eqref{eq:continuity_equation_integrated} are written in terms of the deformed radius $ R(z,t) = 1+\beta  u_{r}(z,t)$, to yield:
\begin{subequations}\begin{align}
\label{eq:flow_rate_defined_r}
    q &= -\frac{R^4}{8}\frac{\partial p}{\partial z},\displaybreak[3]\\
\label{eq:continuity_equation_integrated_r}
   \frac{\partial}{\partial  t} \left( \rho R^2\right) + \frac{\partial (\rho q)}{\partial z} &= 0.
\end{align}\end{subequations}
Substitution of Eq.~\eqref{eq:flow_rate_defined_r} into Eq.~\eqref{eq:continuity_equation_integrated_r} eliminates $q$ and yields:
\begin{equation}
\label{eq:continuity_equation_integrated_r2}
   2\rho  R\frac{\partial  R}{\partial  t} +\frac{\partial \rho}{\partial  t} R^2 - \frac{\partial^2 p}{\partial z^2}\frac{\rho  R^4}{8}-\frac{ R^4}{8}\frac{\partial \rho}{\partial z}\frac{\partial p}{\partial z}-\frac{4 R^3}{8}\frac{\partial  R}{\partial z}\frac{\partial p}{\partial z}\rho=0.
\end{equation}

First, we consider the case of a tube implusively pressurized at the inlet, with its outlet at gauge pressure. So the pressure boundary conditions are:
    \begin{equation}
    \label{eq:Impulsive_Pressure_BC}
    \left. p\right|_{z=0} = H(t),\qquad
    \left. p\right|_{z=1} =0,
\end{equation}
where $H(t)$ is the Heaviside function unit-step function.
Next, a perturbation expansion is introduced in the compressibility parameter\cite{V06,VB10} $\alpha\ll1$:%
\begin{subequations}\label{eq:perturbation_a}\begin{align}
  p &= p^{0}+\alpha p^{1}+\cdots,\\
   u_{r} &=   u_{r}^{0}+\alpha  u_{r}^{1}+\cdots ,\\
   R &=  R^{0}+\alpha R^{ 1}+\cdots, \\
   \rho &= \rho^0 +\alpha\rho^{1}+\cdots \\ 
    &=1+\alpha\left(p^{0}+\alpha p^{1} \cdots \right).
 \end{align}
\end{subequations}
Substituting this perturbation expansion into Eq.~\eqref{eq:continuity_equation_integrated_r2} yields, following tedious but straightforward algebra,\footnote{For completeness, the full expression resulting from substituting the perturbation expansion~\eqref{eq:perturbation_a} into Eq.~\eqref{eq:continuity_equation_integrated_r2} is given as Eq.~\eqref{eq:continuity_equation_integrated_r3} in Appendix~\ref{app:expansions}.} 
\begin{equation}
\label{eq:Fluid_Flow_LeadingOrder}
     2R^{0}\frac{\partial R^{0} }{\partial  t}-\frac{\left( R^{0}\right)^4}{8}\frac{\partial ^2 p^{0}}{\partial z^2}-\frac{1}{2}\left( R^{0}\right)^3\frac{\partial p^{0}}{\partial z}\frac{\partial  R^{0}}{\partial z} =0
\end{equation}
at the leading order, $\mathcal{O}(\alpha^0)$.

The structural mechanics Eq.~\eqref{eq:Deformation_ODE_final} does not explicitly involve $\alpha$, thus at every order of $\alpha^j$ $(j=0,1,\hdots)$ it keeps its form, providing us with
\begin{equation}
\label{eq:Deformation_Equation_expanded}
  u_{r}^{j} + \frac{1}{\De}\frac{\partial u_{r}^{j}}{\partial t} + \St  \frac{\partial^2 u_{r}^{j}}{\partial t^2} =  p^{j}(z) .
\end{equation}

\subsubsection{Leading-order solution}
\label{eq:impulsive_leading_order}

First, we solve the leading-order FSI problem, \textit{i.e.}, Eqs.~\eqref{eq:Fluid_Flow_LeadingOrder} and \eqref{eq:Deformation_Equation_expanded}. Since the problem is still nonlinear, we address the nonlinearity by introducing a (second) perturbation expansion in the small FSI parameter $\beta$: 
\begin{subequations}\label{eq:perturbation_b}\begin{align}
 p^{0} &= p^{0,0}+\beta p^{0,1}+ \cdots,\\
  u_{r}^{0} &=  u_{r}^{0 ,0}+\beta  u_{r}^{0,1} + \cdots,\\
   R^{0} &=  R^{0,0}+\beta R^{0 ,1}+\cdots \\
  &= 1+\beta\left(  u_{r}^{0 ,0}+\beta  u_{r}^{0,1} + \cdots\right).
 \end{align}
\end{subequations}
{Here, the first superscript denotes the perturbation order with respect to $\alpha$, while the second superscript, after the comma, denotes the perturbation order with respect to $\beta$.} Substituting the perturbation expansion~\eqref{eq:perturbation_b} into   Eq.~\eqref{eq:Fluid_Flow_LeadingOrder} yields:
\begin{multline}
\label{eq:fluid_mechanics_first_order}
2\beta \frac{\partial  R^{0 ,1}}{\partial  t}+2\beta^2 R^{0 ,1}\frac{\partial  R^{0 ,1}}{\partial  t}-\frac{1}{8}\left(\frac{\partial ^2p^{0 ,0}}{\partial z^2}+4\beta R^{0 ,1}\frac{\partial ^2 p^{0 ,0}}{\partial z^2}+\beta\frac{\partial ^2 p^{0 ,1}}{\partial z^2}\right) \\
- \frac{\beta}{2} \frac{\partial p^{0 ,0}}{\partial z}\frac{\partial  R^{0 ,1}}{\partial z}+\mathcal{O}(\beta^2) =0.
\end{multline}

To the leading order in $\beta$, we simply have
\begin{equation}
\label{eq:pressure_leading_a_b}
 \frac{\partial ^2p^{0 ,0}}{\partial z^2} =0,
\end{equation}
which, obviously, also holds for the pressure field of an incompressible fluid in a rigid tube ($\alpha=\beta=0$). 
The imposed boundary conditions on the pressure, given by Eq.~\eqref{eq:Impulsive_Pressure_BC}, apply to  $p^{0,0}(z,t)$, while the higher perturbations satisfy the homogeneous boundary conditions. Recall that the tube is impulsively pressurized at the inlet, subject to zero gauge pressure at the outlet, so the boundary conditions for Eq.~\eqref{eq:pressure_leading_a_b} are
\begin{equation}
    \left. p^{0 ,0}\right|_{z=0} = H(t),\qquad
    \left. p^{0 ,0}\right|_{z=1} =0.
\end{equation}
Thus, the leading-order (in both $\alpha$ and $\beta$) solution for an impulsively pressurized tube is simply
\begin{equation}
\label{eq:1_p_{0,0}}
     p^{0 ,0}(z, t) = H( t)(1-z).
\end{equation}

Next, substituting the perturbation expansion from Eq.~\eqref{eq:perturbation_b} into the leading-order-in-$\alpha$ structural mechanical equation~\eqref{eq:Deformation_Equation_expanded} ($j=0$), and collecting the leading-order-in-$\beta$ terms, yields
\begin{equation}
\label{eq:1_Deformation_leading_order_ODE}
\begin{aligned}
u_{r}^{0 ,0} + \frac{1}{\De}\frac{\partial u_{r}^{0 ,0}}{\partial t} + \St\frac{\partial ^2 u_{r}^{0 ,0}}{\partial  t^2} &=   p^{0 ,0}(z) \\ &=  H( t)(1-z) . 
\end{aligned}
\end{equation}
Equation~\eqref{eq:1_Deformation_leading_order_ODE} is subject to the initial conditions corresponding to starting from rest:
\begin{equation}
    u_{r}^{0,0}|_{t=0} = 0,  \qquad
 \left.\frac{\partial  u_{r}^{0,0}}{\partial  t}\right|_{t=0} = 0.
\label{eq:u00_rest_ic}
\end{equation}
Equation~\eqref{eq:1_Deformation_leading_order_ODE} subject Eq.~\eqref{eq:u00_rest_ic} has the following solution:
\begin{equation}
 \label{eq:Leading_Deformation_1}
      u^{0,0}_{r}(z, t) = (1-z) {\Xi}_0(t) ,
\end{equation}
where we have defined
\begin{subequations}
 \begin{align}
  {\Xi}_0( t)  &= H(t)\left\{1 - \left[\cos\left(\Omega t\right) + \frac{1}{\mathfrak{t}_c\Omega}\sin\left(\Omega t\right)\right]\re^{-t/\mathfrak{t}_c}\right\},\\
  \label{eq:Damped_Frequency}
  \Omega &= \sqrt{\frac{1}{\St}-\frac{1}{4\St^2 \De^2}}.
  \end{align}
\end{subequations}
Here, for convenience, we have introduced $\mathfrak{t}_c = {2\St\,\De}$ as a dimensionless time constant. 
Observe that, for the solution from Eq.~\eqref{eq:Leading_Deformation_1} to allow oscillations, the term under the square root in the expression for the damped frequency $\Omega$ from Eq.~\eqref{eq:Damped_Frequency} must be  positive; \textit{i.e.}, the ratio of critical damping $\zeta = 1/(2\De\sqrt{\St}) < 1$.\footnote{The overdamped case of $\zeta > 1$ is not of interest here, as we assume $\De$ is sufficiently large, say $\De>1/(2\sqrt{\St})$, so that the tube's response is primarily elastic, with weaker viscous damping.}

\subsubsection{First-order correction in $\beta$ and $\alpha$}

To obtain the first-order-in-$\beta$ correction to the pressure, namely $p^{0 ,1}(z, t)$, we use Eq.~\eqref{eq:fluid_mechanics_first_order}. Collecting all the terms of $\mathcal{O}({\beta})$, and substituting the expressions for ${\partial u_r^{0,0}}/{\partial z}$, ${\partial u_r^{0,0}}/{\partial t}$ and ${\partial p^{0,0}}/{\partial z}$ calculated from Eqs.~\eqref{eq:Leading_Deformation_1} and \eqref{eq:1_p_{0,0}}, we obtain: 
\begin{equation}
    \label{eq:fluid_mechanics_first_order_b_soln}
     2(1-z){\Xi}_1(t) - \frac{1}{8}\frac{\partial ^2 p^{0 ,1}}{\partial z^2} - \frac{1}{2}{\Xi}_0( t) = 0, \qquad {\Xi}_1(t) = \frac{\partial \Xi_0(t)}{\partial t}.
\end{equation}
The solution of Eq.~\eqref{eq:fluid_mechanics_first_order_b_soln}, subject to homogeneous boundary conditions, is
\begin{equation}
 \label{eq:1_p^{0,1}}
     p^{0,1}(z,t) = 16{\Xi}_1(t)F(z) - 2{\Xi}_0(t)G(z),
\end{equation}
where we have defined
\begin{equation}
F(z) = \frac{z^2}{2}-\frac{z^3}{6}-\frac{z}{3},  \qquad
G(z) = z^2-z.
\label{eq:Auxilliary}
\end{equation}

Next, we solve for the first-order-in-$\alpha$ correction $p^{1,0}$ for the pressure, recalling that the result of substituting the perturbation expansion in $\alpha$ from Eq.~\eqref{eq:perturbation_a} into Eq.~\eqref{eq:continuity_equation_integrated_r2} and collecting terms at $\mathcal{O}(\alpha)$ is provided as Eq.~\eqref{eq:Fluid_Flow_FirstOrder} in Appendix~\ref{app:expansions}. Once again, we now introduce a perturbation expansion in the FSI parameter $\beta$: \begin{subequations}\label{eq:perturbation_b2}\begin{align}
 p^{1} &= p^{1 ,0} + \beta p^{1 ,1} + \cdots,\\
  u_{r}^{1} &=   u_{r}^{1 ,0} + \beta  u_{r}^{1 ,1} + \cdots,\\
   R^{1} &=  R^{1 ,0} + \beta R^{1 ,1} + \cdots,\\
  &=0+\beta\left( u_{r}^{1 ,0} + \beta u_{r}^{1 ,1} + \cdots\right).
 \end{align}
\end{subequations}
We substitute the perturbation expansion~\eqref{eq:perturbation_b2} into the $\mathcal{O}(\alpha)$ equation (\textit{i.e.}, Eq.~\eqref{eq:Fluid_Flow_FirstOrder} in Appendix~\ref{app:expansions}) to obtain\footnote{As before, see Appendix~\ref{app:expansions}, specifically  Eq.~\eqref{eq:Fluid_Flow_FirstOrder_b} for details.}
\begin{equation}
\label{eq:pressure_1_0_differential}
    \frac{\partial p^{0 ,0}}{\partial  t} - \frac{1}{8}\frac{\partial ^2 p^{1 ,0}}{\partial z^2} - \frac{1}{8}\left(\frac{\partial p^{0 ,0}}{\partial z}\right)^2 =0
\end{equation}
at the leading order, $\mathcal{O}(\beta^0)$.
The solution of Eq.~\eqref{eq:pressure_1_0_differential}, subject to homogeneous boundary conditions for $p^{1,0}(z,t)$, is
\begin{equation}
\label{eq:p^{1,0}}
    p^{1 ,0}(z, t) = 8\delta(t)F(z)-\frac{1}{2}H(t)G(z).
\end{equation}

\subsubsection{Summary of the perturbation solution}

In summary, the solution for deformed radius in a viscoelastic tube conveying transient compressible flow with suddenly imposed pressure at the inlet is
\begin{equation}
\label{eq:1_Deformed_Radius_Net}
    {R}({z,t}) = 1+\beta u_r^{0,0}(z,t)
    +\mathcal{O}(\alpha\beta,\beta^2,\alpha^2),
\end{equation}
where $u_r^{0,0}$ is given by Eq.~\eqref{eq:Leading_Deformation_1}. As discussed in Sec.~\ref{sec:vel_field}, $\alpha\ll 1$ and $\beta \ll 1$, therefore all quadratic perturbation terms are neglected. The pressure distribution within the tube is
\begin{equation}
\label{eq:1_Pressure_Net}
    p(z,t) = p^{0,0}(z,t)+\beta p^{0,1}(z,t)+\alpha p^{1,0}(z,t) +\mathcal{O}(\alpha\beta,\beta^2,\alpha^2),
\end{equation}
where $p^{0,0}$ is given by Eq.~\eqref{eq:1_p_{0,0}}, $p^{0,1}$ is given by Eq.~\eqref{eq:1_p^{0,1}}, and $p^{1,0}$ is given by Eq.~\eqref{eq:p^{1,0}}.

\subsection{Oscillatory pressure at the inlet}
\label{sec:oscillatory}
Next, we consider an oscillatory pressure suddenly imposed at the inlet. Now, the boundary condition at the inlet, in dimensional form, is
\begin{equation}
\bar{p}|_{\bar{z} =0} = \mathcal{P}_c H(t)\cos(\bar{\omega}\bar{t}), \qquad
\bar{p}|_{\bar{z} =1} = 0.
\label{eq:boundary_condition_2_1}
\end{equation}
The angular frequency $\bar{\omega}$ imposes a new time scale $\mathcal{T}_o={1}/{\bar{\omega}}$ on the flow. Although, in the nondimensionalization scheme in Sec.~\ref{sec:fluid_cont_eq} we chose $\mathcal{T}_f = \ell/\mathcal{V}_z$, where $\mathcal{V}_z$ was set by $\mathcal{P}_c$ via Eq.~\eqref{eq:velocity_scale}, we could just as well set $\mathcal{T}_f = \mathcal{T}_o$ (while keeping $\mathcal{V}_z$ set by $\mathcal{P}_c$ via Eq.~\eqref{eq:velocity_scale}). Then, the dimensionless governing equations of the flow have the same form as those derived in Sec.~\ref{sec:fluid_mom_eq}, except $\epsilon \Rey$ is replaced\citep{TS57} by $\Wo^2$, where $\Wo$ is the Womersley number, conventionally defined as $\Wo^2 = \rho_0 \bar{\omega} a^2 /\mu$. Consonant with our lubrication assumption that $\epsilon \Rey\ll 1$, $\Wo^2\ll1$ is required. Physically, this assumption means that the time scale of viscous diffusion is much smaller than the time scale of imposed oscillations, and the flow is quasi-static with respect to the forcing \cite{L07}. Therefore, we are justified in keeping our original nondimensionalization in this case.

\subsubsection{Leading-order solution}
Following the double perturbation approach as in Sec.~\ref{sec:impulsive}, the governing equation for $p^{0 ,0}(z, t)$ is still given by Eq.~\eqref{eq:pressure_leading_a_b}. 
Using Eq.~\eqref{eq:nd_vars_tube}, the dimensionless boundary conditions are 
\begin{equation}
    p|_{z=0} = H(t)\cos(\omega t),\qquad
    p|_{z=1} =0,
\label{eq:boundary_condition_2_1_dimless}
\end{equation}
where $\omega = \mathcal{T}_f\bar{\omega}$ is the dimensionless frequency. Thus, $p^{0 ,0}$ must satisfy Eqs.~\eqref{eq:boundary_condition_2_1_dimless} and higher perturbations satisfy homogeneous boundary conditions. Hence,
\begin{equation}
\label{eq:2_p^{0,0}}
    p^{0 ,0}(z, t) = H(t) \cos(\omega t)(1-z).
\end{equation}

Similarly, from Eq.~\eqref{eq:1_Deformation_leading_order_ODE}, the equation governing the leading-order (in $\alpha$ and $\beta$) radial deformation $ u^{0,0}_{r}$ is
\begin{equation}
\label{eq:2_Structural_ODE}
\begin{aligned}
   u_{r}^{0 ,0}+ \frac{1}{\De}\frac{\partial u_{r}^{0 ,0}}{\partial  t} + \St\frac{\partial ^2  u_{r}^{0 ,0}}{\partial  t^2} &=  p^{  0 ,0}(z)\\
     &=  H(t)\cos(\omega t)  (1-z). 
\end{aligned}     
\end{equation}
The solution Eq.~\eqref{eq:2_Structural_ODE}, subject to homogeneous initial conditions (starting from rest) as in Eq.~\eqref{eq:u00_rest_ic}, is 
\begin{equation}
    \label{eq:2_u_r^{0,0}}
    u^{0 ,0}_{r}(z, t) = (1-z)\Psi_0( \omega,t), 
\end{equation}
where we have defined
\begin{subequations}\begin{align}
    \Psi_0( \omega,t) &= H(t)\left[\hat{A}(\omega)\cos(\omega t)+\frac{\hat{B}(\omega)}{\omega}\sin(\omega t) \right]\\
    &\phantom{=}
    +H(t)\left\{-\hat{A}(\omega)\cos\left(\Omega t\right)  +\left[\frac{\hat{C}(\omega)}{\St\,\Omega} +\frac{\hat{A}(\omega)}{\mathfrak{t}_c\Omega}\right]\sin\left(\Omega t\right)\right\}\re^{-t/\mathfrak{t}_c},\nonumber\\
    \label{eq:A_Hat}
    \hat{A}(\omega) &=\frac{\left(1-\St\,\omega^2 \right) \De^2}{\left(\De-\St\,\De\, \omega^2\right)^2+\omega^2}, \\
    \label{eq:B_Hat}
    \hat{B}(\omega) &=\frac{\De\,\omega^2}{\left(\De-\St\, \De\, \omega^2\right)^2+\omega^2}, \\
    \label{eq:C_Hat}
    \hat{C}(\omega) &=-\frac{\De}{\left(\De-\St\,\De\,\omega^2\right)^2+\omega^2}.
\end{align}\end{subequations}
As before, $\mathfrak{t}_c = 2\St\, \De$, and $\Omega$ is given by Eq.~\eqref{eq:Damped_Frequency}.

\subsubsection{First-order corrections in $\beta$ and $\alpha$}
The governing equation for $p^{0 ,1}({z,t)}$ is obtained as in Sec.~\ref{sec:impulsive} (recall the  derivation of Eq.~\eqref{eq:fluid_mechanics_first_order_b_soln}), and its solution, subject to homogeneous boundary conditions, is
\begin{equation}
\label{eq:2_p^{0,1}}
    p^{0,1}(z, t) = {16}F(z){\Psi}_1(\omega, t) - 2 G(z){\Psi}_0(\omega, t) \cos(\omega t), \qquad \Psi_1 (\omega ,t) = \frac{\partial \Psi_0 (\omega ,t)}{\partial t}.
\end{equation}
Similarly, $p^{1,0}$ is still governed by Eq.~\eqref{eq:pressure_1_0_differential}, and its solution, given the leading-order solutions in Eqs.~\eqref{eq:2_p^{0,0}} and \eqref{eq:2_u_r^{0,0}}, is
\begin{equation}
\label{eq:2_p^{1,0}}
    p^{1,0}(z, t) = F(z)[8\delta(t)\cos(\omega t) - 8\omega H(t) \sin(\omega  t)] - \frac{1}{2}G(z) H(t) \cos^2(\omega t).
\end{equation}

\subsubsection{Summary of the perturbation solution}

In summary, the deformed radius of a viscoelastic tube conveying transient flow with oscillatory pressure imposed at the inlet is:
\begin{equation}
\label{eq:2_Deformed_Radius_Net}
    {R}(z,t) = 1+\beta u_r^{0,0}(z,t)
    +\mathcal{O}(\alpha\beta,\beta^2,\alpha^2),
\end{equation}
where as before, all quadratic perturbation terms are neglected.
Here, $u_r^{0,0}$ is given by Eq.~\eqref{eq:2_u_r^{0,0}}. The expression for the pressure  distribution within the tube is
\begin{equation}
\label{eq:2_Pressure_Net}
    p(z,t) = p^{0,0}(z,t)+\beta p^{0,1}(z,t)+\alpha p^{1,0}(z,t) +\mathcal{O}(\alpha\beta,\beta^2,\alpha^2).
\end{equation}
Here, $p^{0,0}$ is given by Eq.~\eqref{eq:2_p^{0,0}} and $p^{0,1}$ is given by Eq.~\eqref{eq:2_p^{0,1}} and $p^{1,0}$ is given by Eq.~\eqref{eq:2_p^{1,0}}.

\section{Results and discussion}
\label{sec:results_discussion}

Having developed a mathematical theory of transient FSIs caused by compressible flow in a viscoelastic tube (Sec.~\ref{sec:problem_formulation}), as well as  perturbative solutions to steady (Sec.~\ref{sec:steady_state_math}) and unsteady problems (Sec.~\ref{sec:impulsive} and Sec.~\ref{sec:oscillatory}), in several distinguished limits, we use these results to discuss the physics of the FSI response in this section.

\subsection{Steady response}
In this subsection, we discuss the characteristics of steady compressible flow  in a linearly elastic tube, based on the analysis in Sec.~\ref{sec:steady_state_math}. 

\begin{figure}[t]
\centering
\subfloat[]{\includegraphics[width=0.45\linewidth]{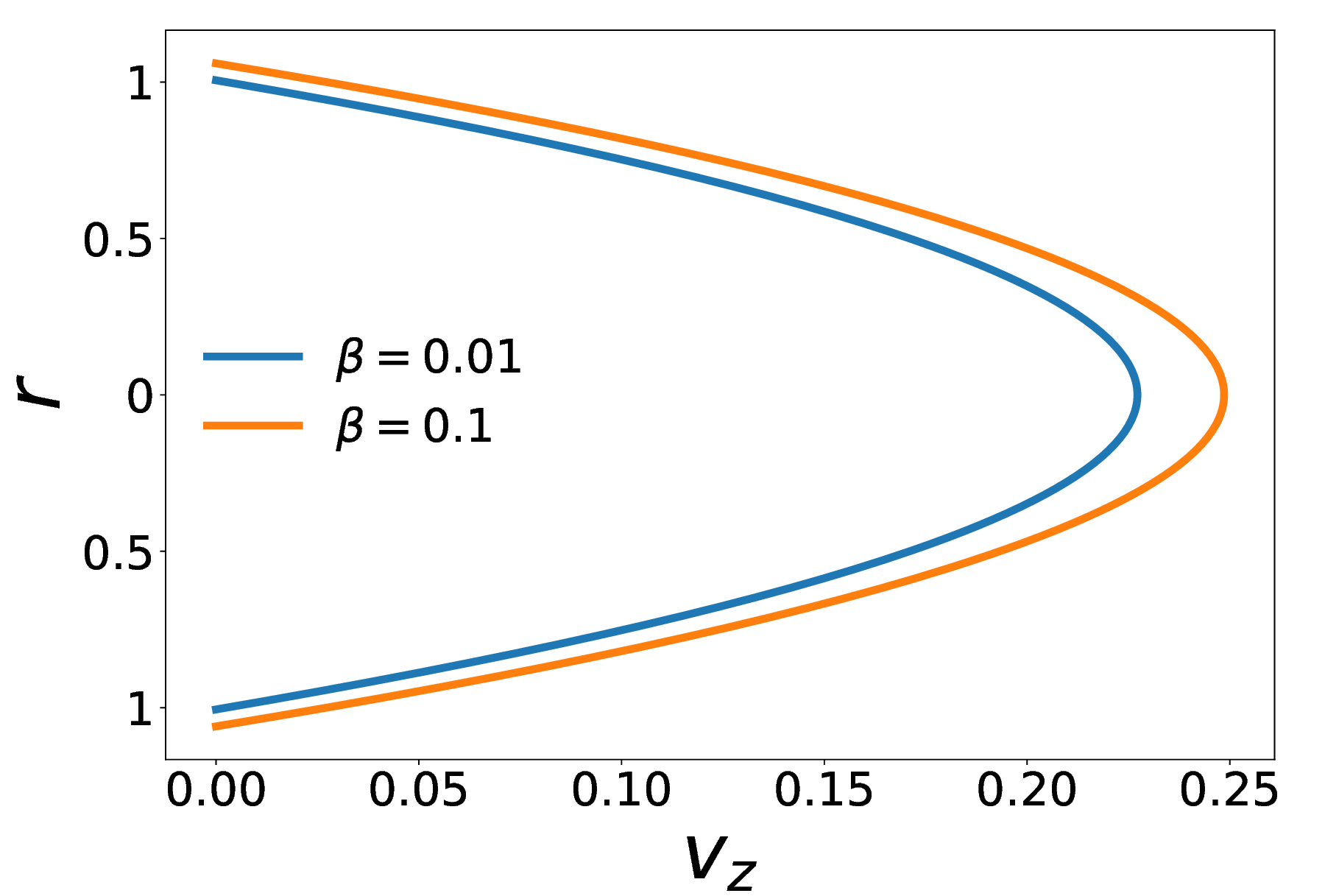}}
\hfill
\subfloat[]{\includegraphics[width=0.45\linewidth]{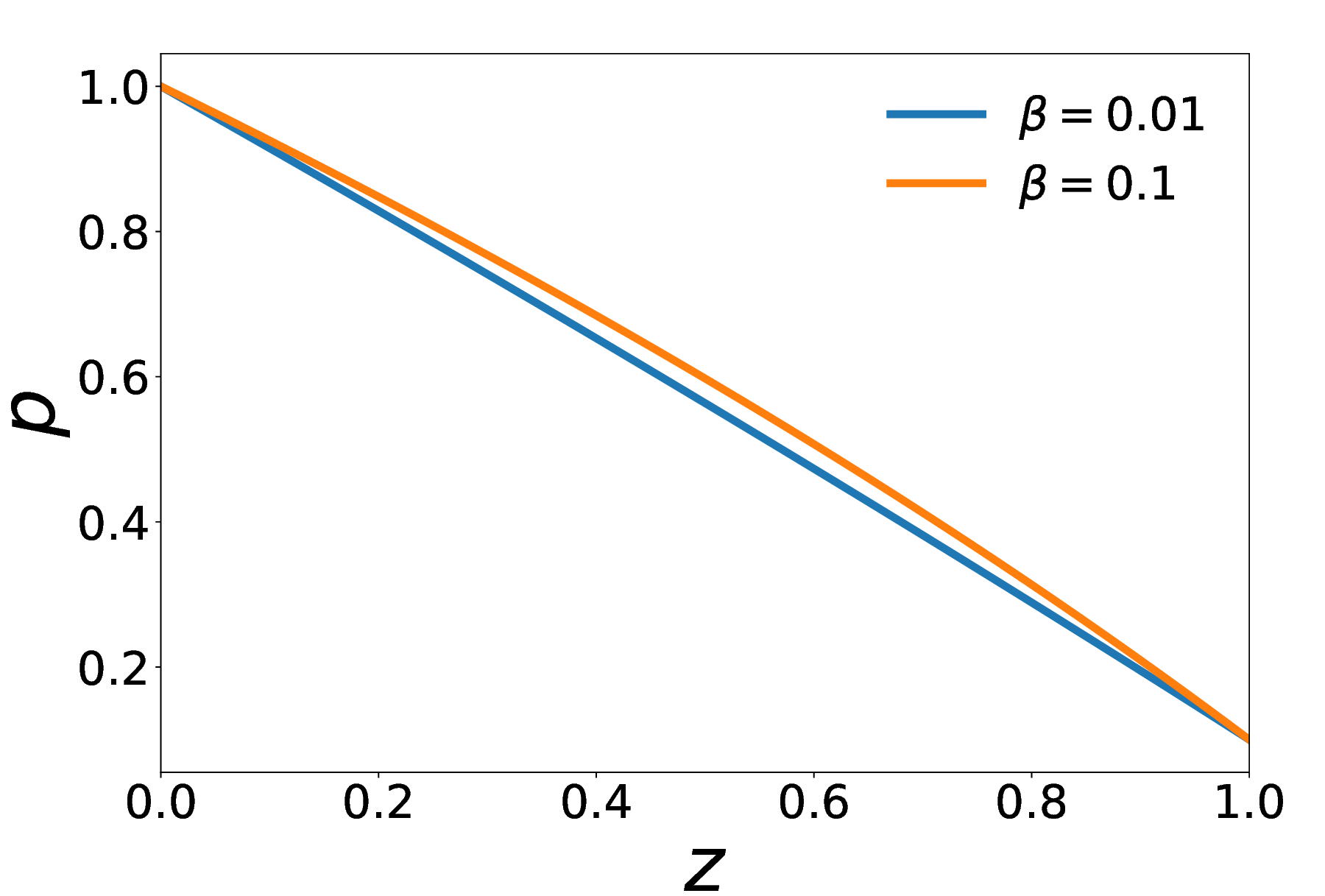}}
\caption{Steady state: $\beta$ dependence of (a) the axial velocity profile $v_z(r,z)$ at $z=0.5$, and (b) the pressure distribution $p(z)$, obtained by inverting Eq.~\eqref{eq:pressure_steady_general}. In both panels: $ \alpha =  0.1$.}
\label{fig:Steady_beta}
\end{figure}

First, we explore the effect of FSI on the flow field through the dimensionless group $\beta$. Higher values of $\beta$ correspond to a wider tube and, consequently, higher throughput (volumetric flow rate $q$), which results in an enhanced average axial velocity (consequently, larger centerline velocity), as shown by the profiles in  Fig.~\ref{fig:Steady_beta}(a). Thus, the fluid exerts more pressure (for larger $\beta$) on the tube wall, on average, as supported by the example plot in Fig.~\ref{fig:Steady_beta}(b).

\begin{figure}[t]
\centering
\subfloat[]{\includegraphics[width=0.45\linewidth]{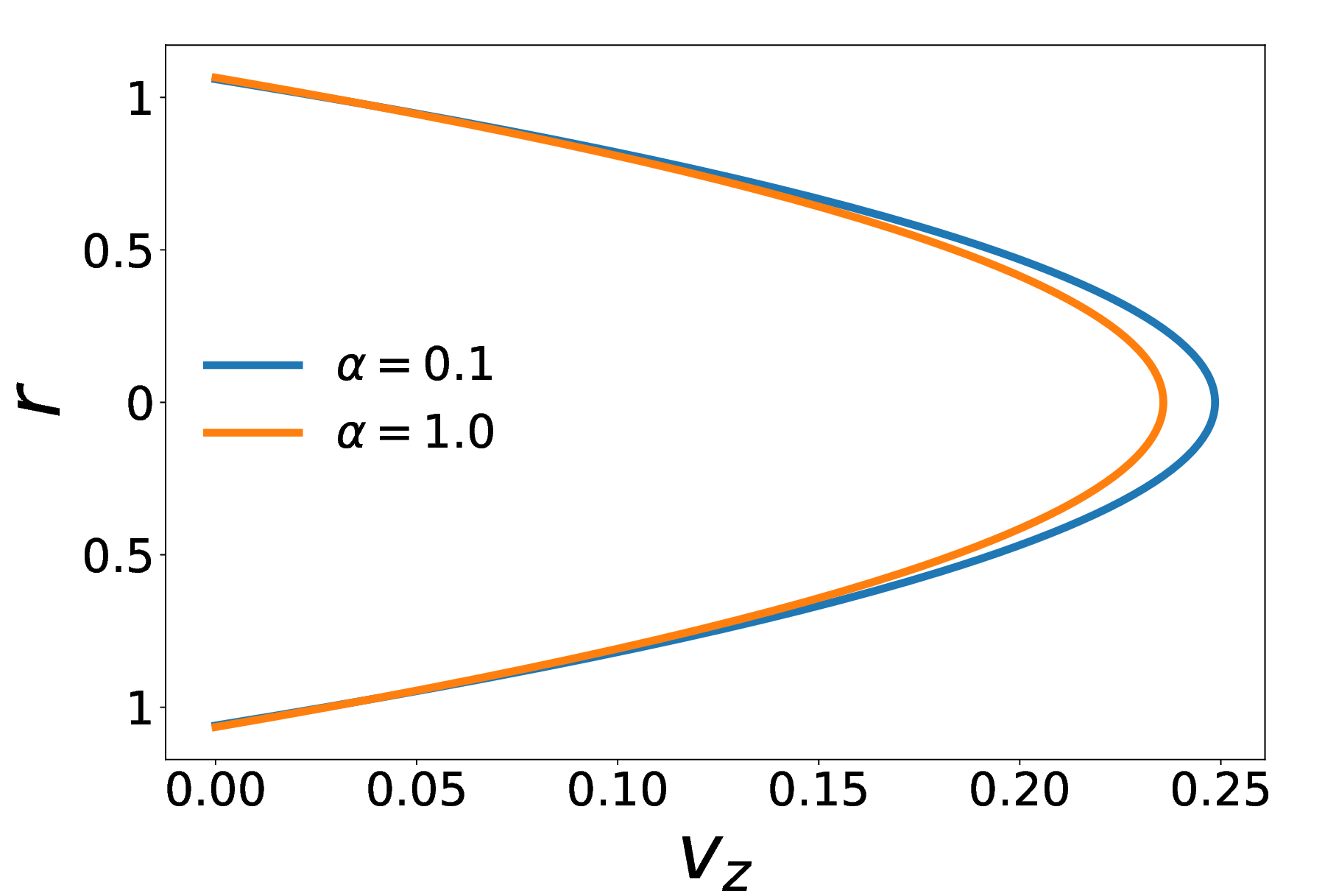}}
\hfill
\subfloat[]{\includegraphics[width=0.45\linewidth]{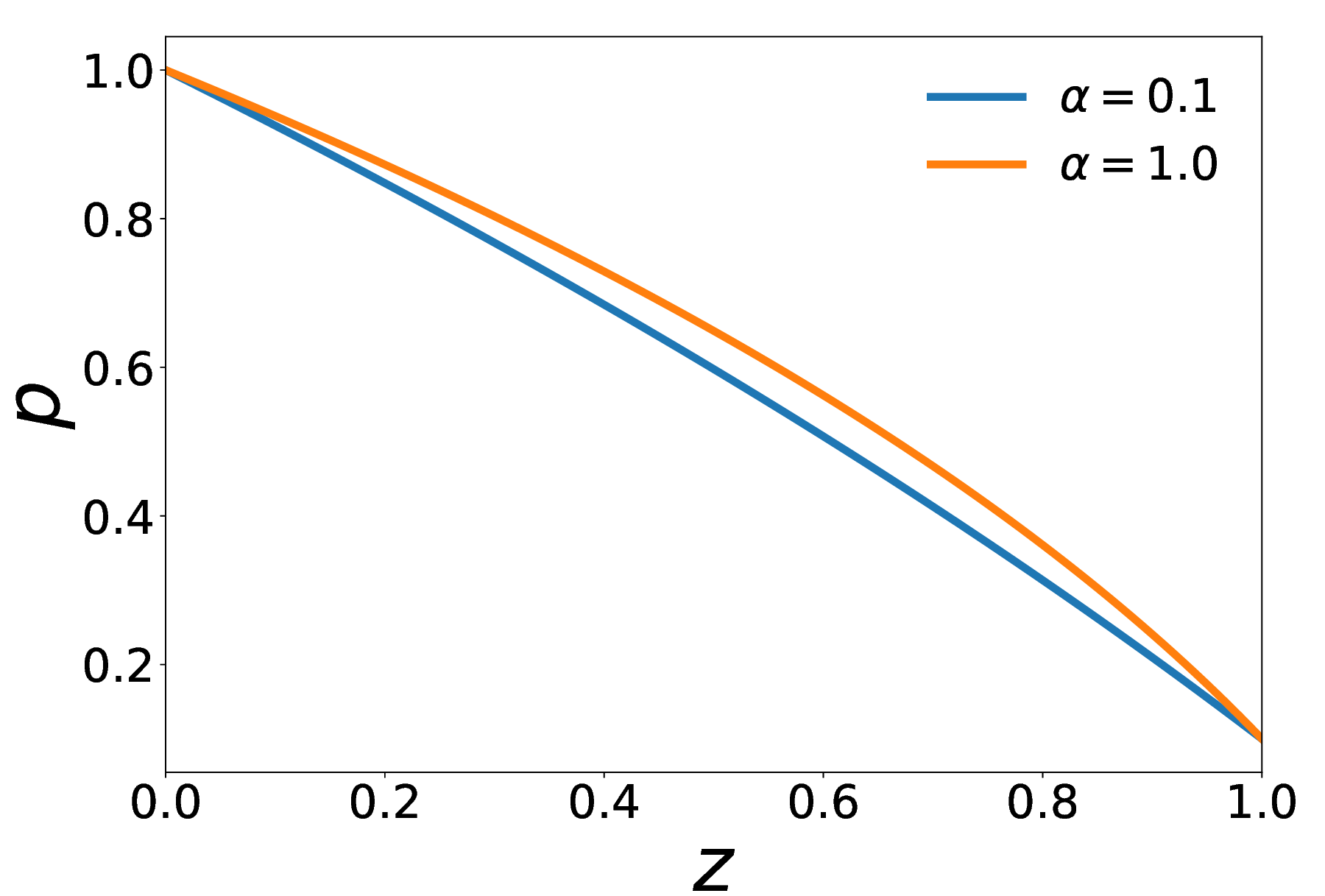}}
\caption{Steady state: $\alpha$ dependence of (a) the axial velocity profile $v_z(r,z)$ at $z =0.5$, and (b) the pressure distribution $p(z)$  obtained by inverting Eq.~\eqref{eq:pressure_steady_general}. In both panels: $\beta =  0.1$}
\label{fig:Steady_A}
\end{figure}

Meanwhile, the effect of the compressibility parameter $\alpha$ on the flow field is shown in Fig.~\ref{fig:Steady_A}. As explained earlier, the compressibility parameter $\alpha$ quantifies the ``sensitivity'' of fluid's density to the applied pressure. With all other parameters fixed, a larger value of $\alpha$ means a larger $\rho$ locally. Therefore, for the same pressure difference, the maximum velocity is reduced by an increase in $\alpha$, as observed in Fig.~\ref{fig:Steady_A}(a). Similarly, since pumping a denser fluid requires higher pressure locally (keeping all other parameters fixed), the pressure inside the tube increases with $\alpha$, as observed in Fig.~\ref{fig:Steady_A}(b).

\subsection{Impulsive pressurization of the inlet}
\label{sec:result_impulsive}

\subsubsection{Deformation and pressure response}
For the case of a viscoelastic tube impulsively pressurized at the inlet, the deformed radius is given by Eq.~\eqref{eq:1_Deformed_Radius_Net} and the pressure profile is given by Eq.~\eqref{eq:1_Pressure_Net}, along with the pertinent supporting equations.  From Eqs.~\eqref{eq:1_Deformed_Radius_Net} and  \eqref{eq:1_Pressure_Net}, we deduce that $R(z,t)$ and $p(z,t)$ have a decaying transient on top of a steady state. The steady state of the deformed radius is given by
\begin{equation}
\label{eq:1_R_steady}
    \lim_{t\to\infty} R(z,t) 
    = 1+\beta (1-z) + \mathcal{O}(\alpha\beta,\alpha^2,\beta^2).
\end{equation}
The pressure profile at the steady state is given by
\begin{equation}
\label{eq:1_p_steady}
    \lim_{t\to\infty} p(z,t) = (1-z) \left(1 + 2\beta z + \frac{\alpha}{2}z \right) + \mathcal{O}(\alpha\beta,\alpha^2,\beta^2).
\end{equation}

\begin{figure}[t]
\centering
\subfloat[Deformed tube radius at $z=0.5$]{\includegraphics[width=0.45\linewidth]{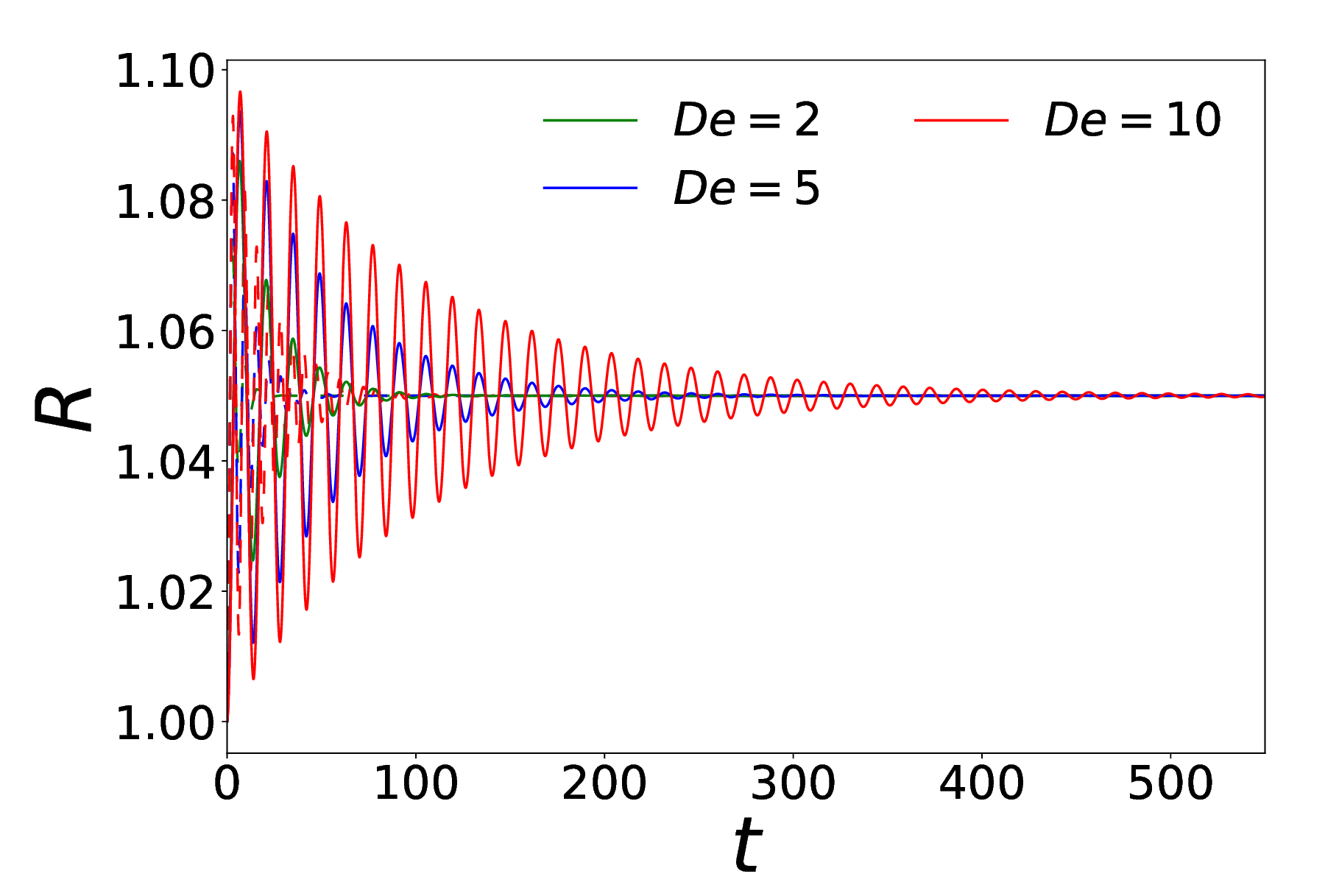}}
\hfill
\subfloat[Pressure at $z=0.5$]{\includegraphics[width=0.45\linewidth]{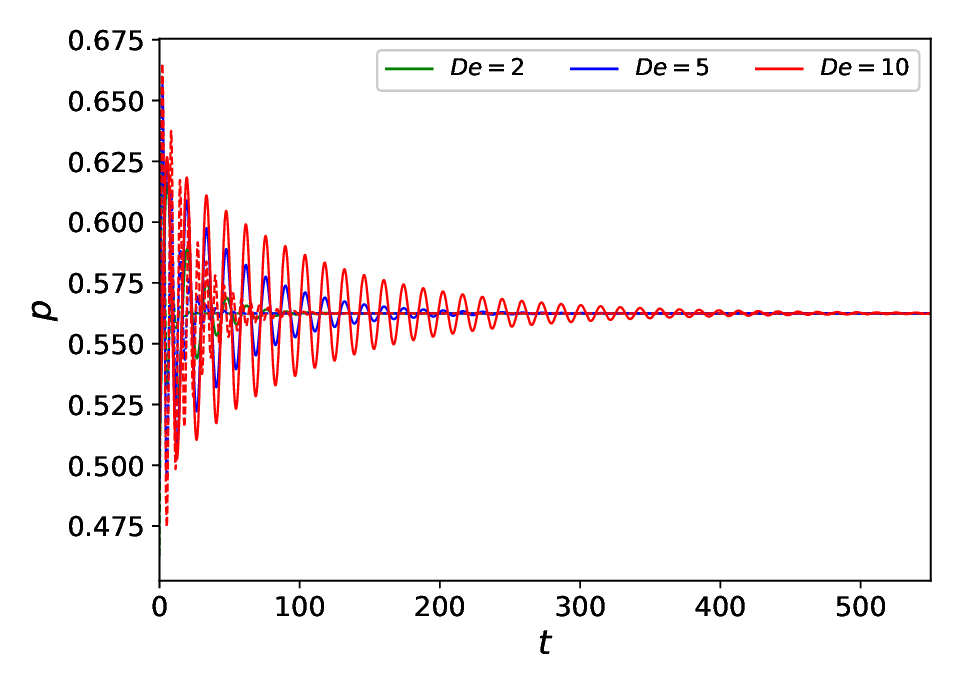}}
\caption{Evolution of (a) the deformed tube radius $R$ and (b) pressure $p$, both at the midlength section ($z =0.5$), for different values of $\De$ and $\St$ in the case of an impulsively pressurized tube, obtained from Eqs.~\eqref{eq:1_Deformed_Radius_Net}--\eqref{eq:1_Pressure_Net}. The solid curves correspond to $\St= 5$, while the dashed curves correspond to $\St= 1$.}
\label{fig:1_R_St_De}
\end{figure}

\begin{figure}[ht]
\centering
\subfloat[Deformed radius at $z=0.5$]{\includegraphics[width=0.45\linewidth]{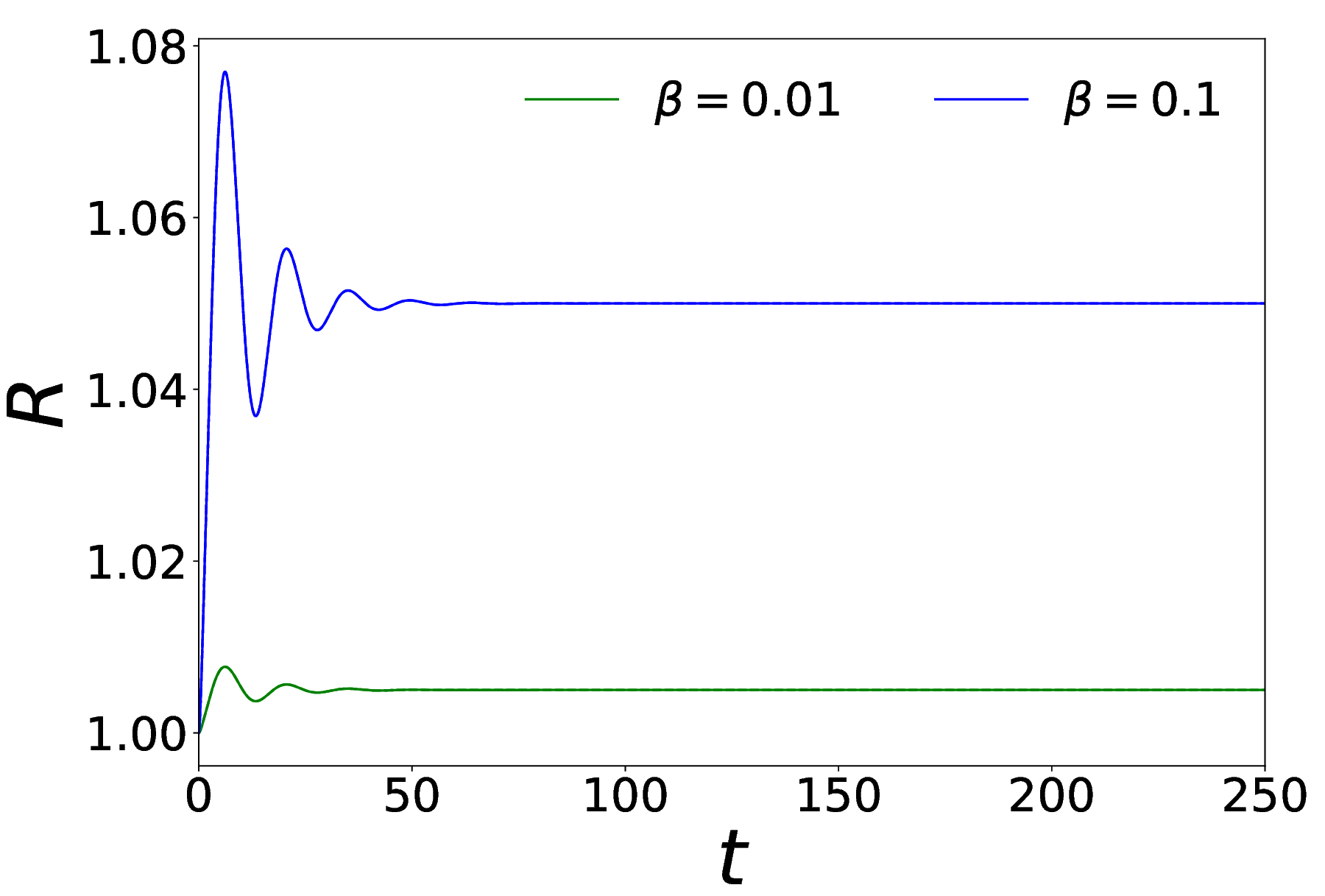}}
\hfill
\subfloat[Pressure at $z=0.5$]{\includegraphics[width=0.45\linewidth]{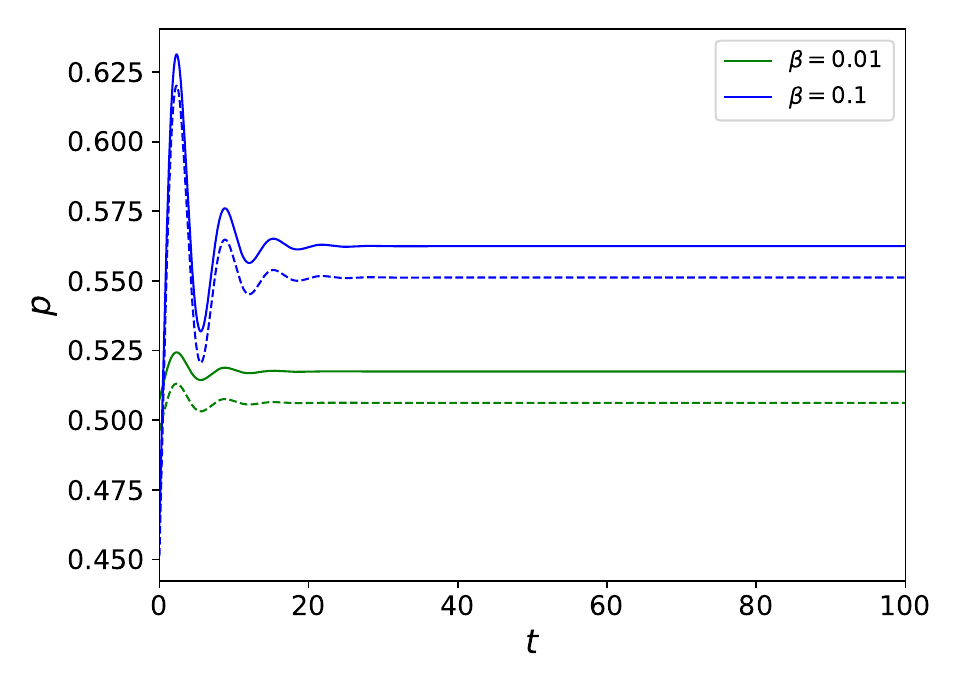}}
\caption{Evolution of (a) the deformed tube radius $R$ and (b) pressure $p$, both at the midlength section ($z =0.5$) for different values of $\beta$ and $\alpha$, for the case of impulsively pressurized tube, as obtained from Eqs.~\eqref{eq:1_Deformed_Radius_Net}--\eqref{eq:1_Pressure_Net}.  The solid curves correspond to $\alpha = 0.1$, while the dashed curves correspond to $\alpha = 0.01$. In both panels: $\De= 2$ and $ \St= 1$.}
\label{fig:1_R_A_Beta}
\end{figure}

The time evolution of $R$ and $p$ are shown in panels (a) and (b), respectively, of Figs.~\ref{fig:1_R_St_De} and \ref{fig:1_R_A_Beta}. Obviously, the steady states are independent of $\De$ and $\St$. The nonzero constant deformation at steady state, after a suddenly applied pressure load, is the result of the bounded \emph{creep} response of the chosen KV model \citep{F93}. That is, the steady state is characterized only by the elasticity (not viscoelasticity) of the system, and the ratio of applied load to the elastic constant determines the local deformation at steady state, which in this case is found from Eq.~\eqref{eq:1_Deformation_leading_order_ODE} to be $=1$. In other words,
\begin{equation}
    \label{eq:1_u_r_0_0_steady}
   \lim_{t\to\infty} u_r^{0,0}(z,t) 
     =  \lim_{t\to\infty} p^{0,0}(z,t) = 1-z.
\end{equation}
Once the transients die out, the inertial and viscoelastic effects are gone, and, naturally, Eq.~\eqref{eq:1_u_r_0_0_steady} agrees with the steady-state result for a linearly elastic tube (without inertia) \citep{AC18b}. 

Next, we focus on the relationship between the transient characteristics of the system and $\St$ and $\De$. We recall that $\St$ quantifies the inertial response of the tube, while $\De$ quantifies the damping (dissipation) of the viscoelastic structure. Therefore, an increase in either $\St$ or $\De$ would prolong the transient response of the tube, as these higher values correspond to tubes with enhanced capacity to store energy, as opposed to  capacity to dissipate energy. Indeed, the transient part of $u_r^{0,0}$ from Eq.~\eqref{eq:Leading_Deformation_1} is
\begin{equation}
 \label{eq:Leading_Deformation_1_transient2}
     \mathfrak{U}_r(z,t) = (1-z)\left[-\cos\left(\Omega t\right) - \frac{1}{\mathfrak{t}_c\Omega}\sin\left(\Omega t\right)\right]\re^{-t/\mathfrak{t}_c} \qquad (t > 0),
\end{equation}
where recall that $\mathfrak{t}_c = {2\St\,\De}$ is a dimensionless time constant, and $\Omega$ is the damped frequency of the system, given by  Eq.~\eqref{eq:Damped_Frequency}. Now, the straightforward observation about the effect of $\St$ and $\De$ on the transient is obvious and also corroborated by Fig.~\ref{fig:1_R_St_De}.

We observe from the plots in Fig.~\ref{fig:1_R_St_De} that both $\St$ and $\De$ also enhance the amplitude of the oscillations. This is again attributed to the enhanced energy storing capacity of the tube at higher values of $\St$ and $\De$. To quantify this effect, we express the damped frequency as
\begin{equation}
\label{eq:damped_frequncy_natural}
    \Omega = \omega_{o}\sqrt{1-\zeta^2},
\end{equation}
where $\omega_{o} = \sqrt{1/\St}$ is the ``natural'' frequency of the system (without any damping) and $\zeta = 1/(2\De\sqrt{\St})$ is  the critical damping ratio of the system, which controls the ratio of successive peaks at a location $z$ in $\mathfrak{U}_r(z, t)$. A fluid particle reaches its consecutive maximum (or minimum)  location after a time period $T = 2\pi/\Omega$. So, we calculate the ratio of the successive peaks in the transient response at a given location and separated by one period, which may be construed as a form of decay rate,  to be
\begin{equation}
\label{eq:amplitude_ratio}
    \frac{\mathfrak{U}_r(z, t+T)}{\mathfrak{U}_r(z, t)}  = \exp\left(-\frac{2\pi\zeta}{\sqrt{1-\zeta^2}}\right).
\end{equation}
From Eq.~\eqref{eq:amplitude_ratio}, it follows that both $\De$ and $\St$ influence the decay rate, through $\zeta$.

It is also instructive to consider the transient response of the pressure field, which is found from Eq.~\eqref{eq:1_p^{0,1}} to be 
\begin{equation}
  \label{eq:pressure_transient}
    \mathfrak{P}(z,t) = \beta\Bigg\{ 16\left(
    \Omega + \frac{1}{\mathfrak{t}_c^2\Omega} \right)\sin\left(\Omega t\right) F(z)
    +2\Bigg[\cos\left(\Omega t\right) + \frac{1}{\mathfrak{t}_c\Omega}\sin\left(\Omega t\right)\Bigg]G(z)\Bigg\} \re^{- t/\mathfrak{t}_c} \quad (t >0).
\end{equation}
First, note that $\mathfrak{P}(z,t)=\mathcal{O}(\beta)$, thus this transience is solely due to FSI, and compressibility has no effect on the transient pressure response. To explain this observation, we note from Eq.~\eqref{eq:z_momentum_dimless2} that the variable density (compressibility) influences the local acceleration of the flow field. However, due to the smallness of reduced Reynolds number $\epsilon \Rey$, the local acceleration of the flow field has been neglected, and therefore compressibility does not affect the transient pressure field. 

On the other hand, Stokes flow in a rigid conduit is inertialess and reacts instantaneously to any unsteadiness imposed by its boundaries \citep{panton}. However, as Eq.~\eqref{eq:pressure_transient} shows, FSI introduces a delay in the response of the fluid by perpetuating exponentially decaying transients in the flow. Figure~\ref{fig:1_R_A_Beta} highlights the FSI-induced transients, showing that the time taken by the system (both the deformed radius and the pressure) to equilibrate increases with $\beta$. Of course, this equilibration time is independent of $\alpha$ (compare the solid and dashed curves). 

\subsubsection{Volumetric flow rate enhancement}

The expression for the volumetric flow rate is  found from Eq.~\eqref{eq:flow_rate_defined}:
\begin{equation}
 \label{eq:flow_rate_no_slip}
      q(z,t) =-\frac{1}{8}\frac{\partial p}{\partial z}\big[1+4\beta u_r(z,t)\big] + \mathcal{O}(\beta^2).
\end{equation}
 For post-transient steady flow in an impulsively pressurized tube, we substitute the expressions from Eqs.~\eqref{eq:1_R_steady} and \eqref{eq:1_p_steady} into Eq.~\eqref{eq:flow_rate_no_slip} to obtain:
\begin{equation}
    q(z) =\underbrace{-\frac{1}{8}\frac{\partial p^{0,0}}{\partial z}}_{q^{0,0}}+\frac{\beta}{4} G'(z) +\frac{\alpha}{16}G'(z)\\-\frac{\beta}{2}(1-z) \frac{\partial p^{0,0}}{\partial z} 
    +\mathcal{O}(\alpha\beta,\alpha^2,\beta^2).
\end{equation}
Here, $q^{0,0}$ denotes the volumetric flow rate in absence of FSI and compressibility. Therefore the ``volumetric flow rate enhancement" due to FSI and compressibility is given by
\begin{equation}
\label{eq:2_enhancement_in_flow_rate}
\begin{aligned}
    q^*(z) = q(z)-q^{0,0}(z) &= \frac{\beta}{4} G'(z) +\frac{\alpha}{16}G'(z) - \frac{\beta}{2}(1-z) \frac{\partial p^{0,0}}{\partial z} \\
     &= \frac{\beta}{4}+\frac{\alpha}{16} \left(2z-1\right).
\end{aligned}     
\end{equation}

\begin{figure}[ht]
\centering
\subfloat[$\beta =0.01$ ]{\includegraphics[width=0.45\linewidth]{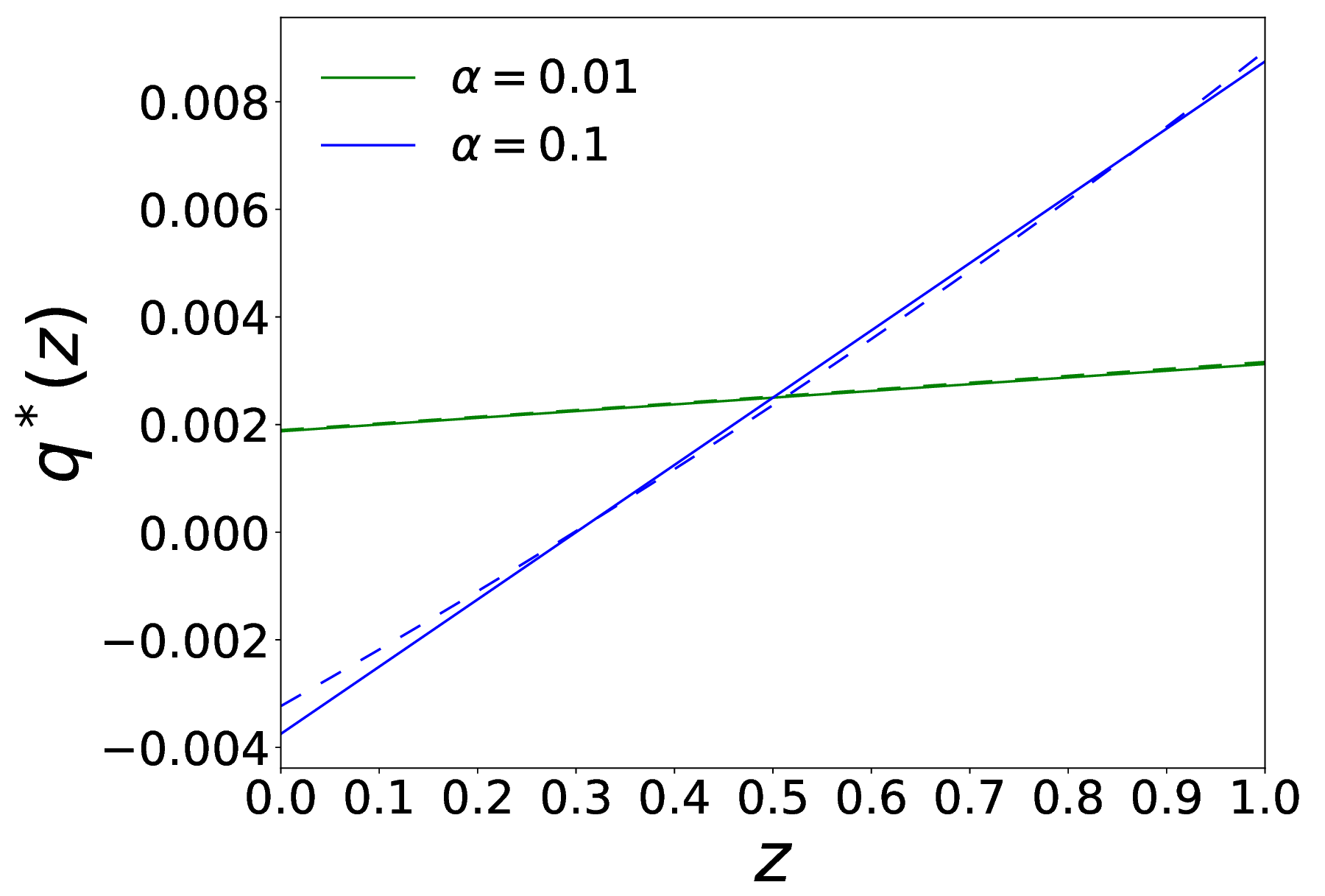}}
\hfill
\subfloat[$\alpha = 0.1$ ]{\includegraphics[width=0.45\linewidth]{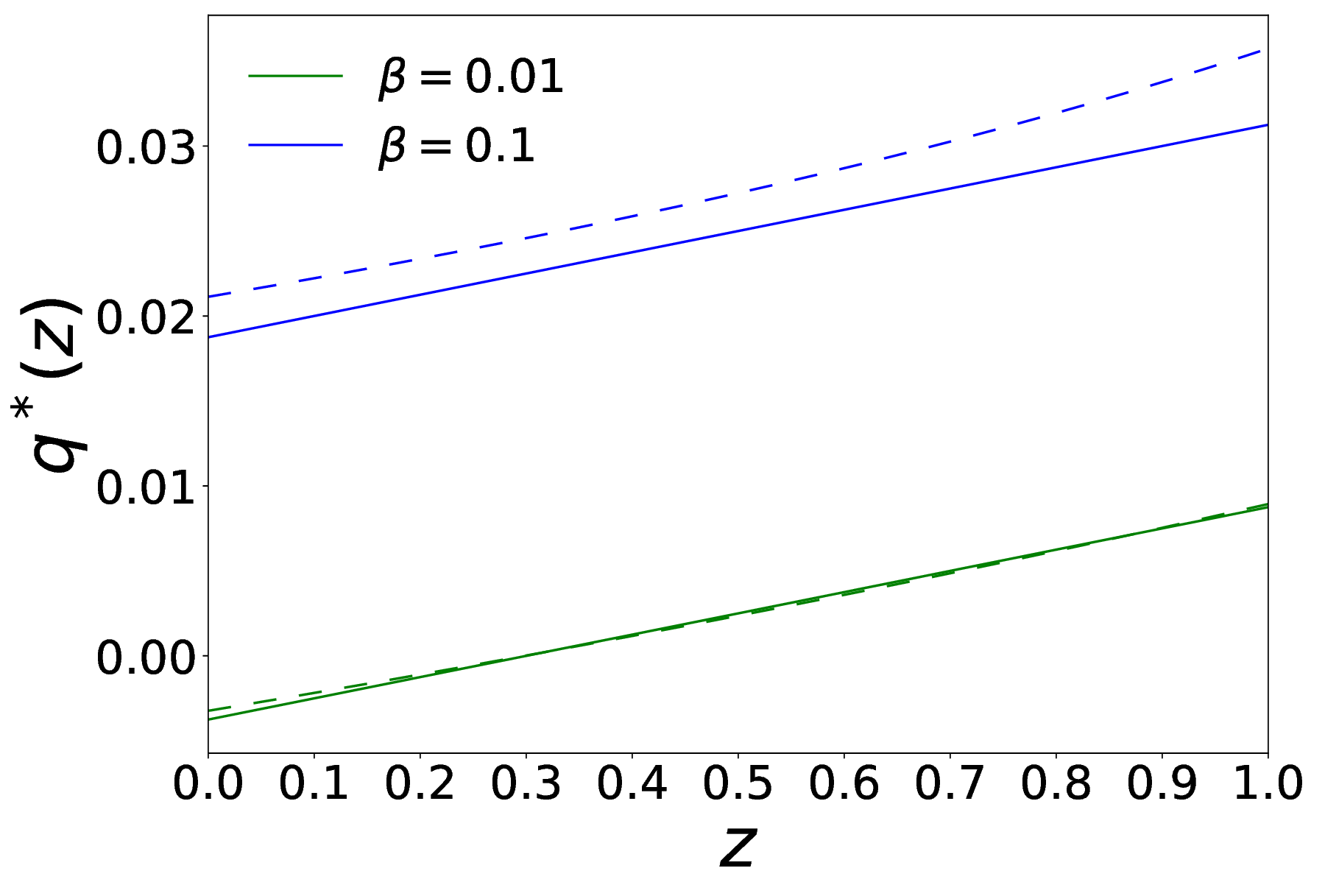}}
\caption{The profile of enhancement in volumetric flow rate $q^*(z)$ for different values of (a) $\alpha $ and (b) $\beta $. The solid curves correspond to the post transient response of the impulsively pressurized tube, solved via perturbation techniques, \textit{i.e.}, Eq.~\eqref{eq:2_enhancement_in_flow_rate}. The dashed curves are the steady-state solution obtained from by numerically inverting Eq.~\eqref{eq:pressure_steady_general}.}
\label{fig:1_Enhancement_flow_Rate_A_Beta}
\end{figure}

Note that even though the flow is steady, both $q$ and $q^*$ (shown in  Fig.~\ref{fig:1_Enhancement_flow_Rate_A_Beta}) still vary along the axial $z$-direction due to  compressibility. For weak compressibility ($\alpha\ll1$), the variation of $q^*$ across the tube is quite weak (see the lighter curves in Fig.~\ref{fig:1_Enhancement_flow_Rate_A_Beta}(a)). Note that $q^{*}(0) = \beta/4 - \alpha/16$ so that $q^*(0) < 0$ for $\beta < \alpha/4$. Meanwhile $q^{*}(1) = \beta/4 + \alpha/16 > 0$ for all $\alpha$ and $\beta$. The strictly increasing nature of $q^*$ with $z$ follows from the fact that the pressure gradient in the tube gives rise to a density gradient, and therefore the density decreases along the tube. Finally, for comparison, we have also plotted the enhancement in flow rate obtained from the steady state analysis of Sec.~\ref{sec:steady_state_math} (dashed curves in Fig.~\ref{fig:1_Enhancement_flow_Rate_A_Beta}). It is clear that the steady-state solution, obtained  numerically (without restrictions on the magnitudes of $\alpha$ or $\beta$), follows closely the post-transient solution obtained via the perturbation expansion, especially for $\alpha,\beta\ll1$, as should be expected.

The enhanced mass flow rate at the outlet is simply $ \dot{m}^*(1) \equiv \left(\rho q^*\right)|_{z=1} = \beta/4 + \alpha/16$ since $\rho(1)=1$ in our nondimensionalization. In the post-transient FSI regime, the mass flow rate is constant, independent of $z$, and equal to $\dot{m}^*(1)$  throughout the tube.

\subsection{Oscillating pressure at the inlet}
\label{sec:result_oscillatory}

Next, we analyze the FSI due to an oscillating pressure imposed at the tube's inlet. For this boundary condition, the solution for the deformed radius $R(z,t)$ is given by  Eq.~\eqref{eq:2_Deformed_Radius_Net}, while the pressure profile $p(z,t)$ is given by Eq.~\eqref{eq:2_Pressure_Net}. 

\subsubsection{Deformation response}
As before, $R(z,t)$ consists of an exponentially decaying transient and a post-transient (\textit{i.e.}, quasi-steady) component. The exponentially decaying transient has the same time constant $\mathfrak{t}_c$ as the impulsively pressurized tube (Sec.~\ref{sec:result_impulsive}), and the same oscillation frequency $\Omega$ given in Eq.~\eqref{eq:damped_frequncy_natural}. Therefore, in this section, we concern ourselves only with the post-transient response of the system, which has the same frequency $\omega$ as the forcing frequency. 

After the transients die out, the post-transient deformed radius is given by
\begin{equation}
 \label{eq:2_R_Steady_1}
    \mathscr{R}(z,t) = 1 +\beta \mathscr{U}_r^{0,0}(z,t) + \mathcal{O}(\alpha\beta,\alpha^2,\beta^2),
 \end{equation}
where
\begin{equation}
 \label{eq:2_R_Steady_2}
   \mathscr{U}_r^{0,0}(z,t)  =  \left[ \hat{A}(\omega)\cos(\omega t)+\frac{\hat{B}(\omega)}{\omega}\sin(\omega t) \right](1-z).
\end{equation}
On inserting the relevant expressions for $\hat{A}$ and $\hat{B}$ from Eqs.~\eqref{eq:A_Hat} and \eqref{eq:B_Hat} into the Eqs.~\eqref{eq:2_R_Steady_1} and \eqref{eq:2_R_Steady_2} and simplifying,  the post-transient deformed radius is found to be:
\begin{equation}
    \mathscr{R}(z,t) = 1 + \beta\frac{\De\,(1-z)}{\sqrt{\left(1-\St\,\omega^2\right)^2\De^2+{\omega^2}}}\sin{(\omega t +\phi)}, \qquad \phi =\tan^{-1}\left(\frac{\hat{A}(\omega)\omega}{\hat{B}(\omega)}\right).
\label{eq:2_R_steady_6}
\end{equation}

\begin{figure}
\centering
\subfloat[Standing wave]{\includegraphics[width=0.475\linewidth]{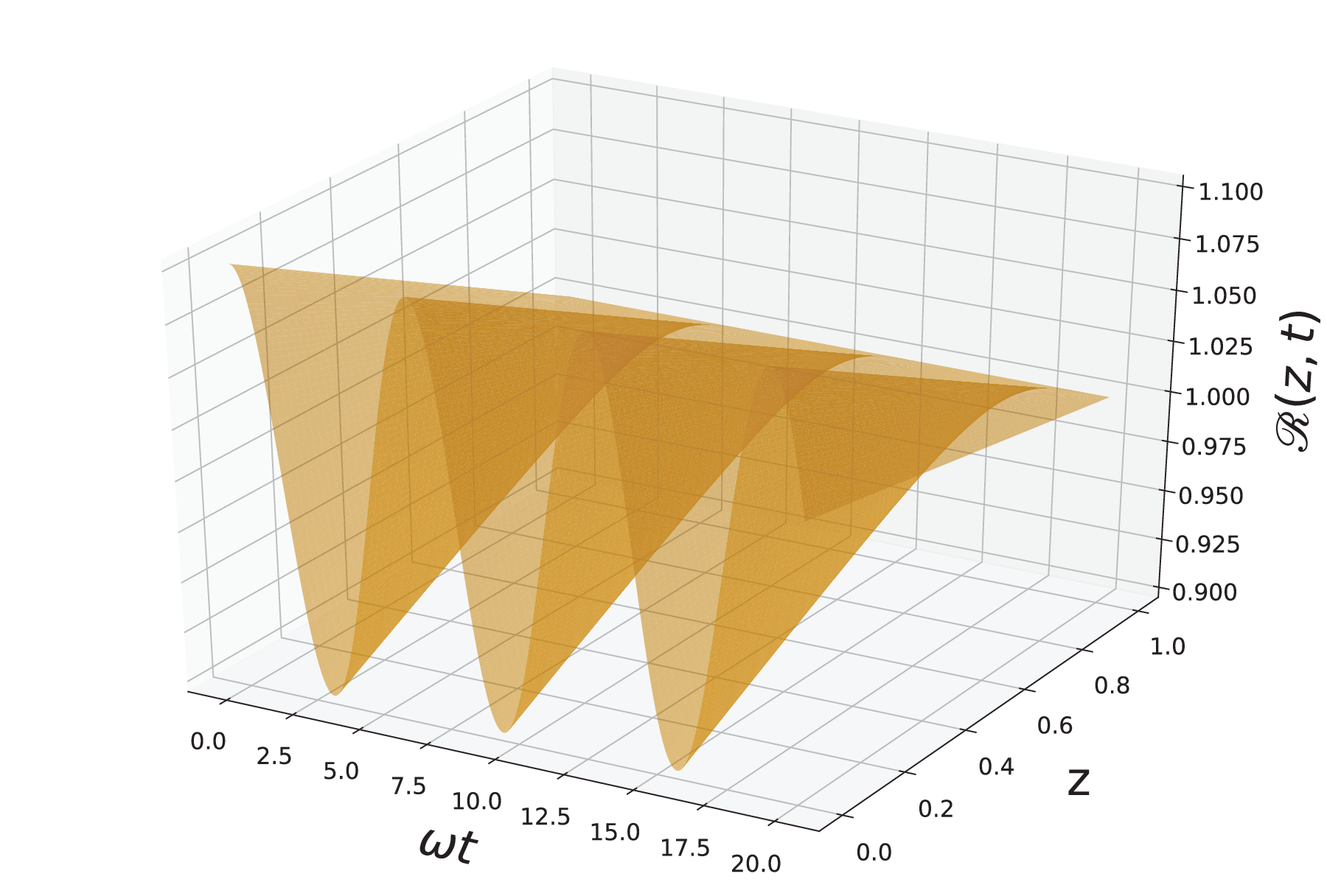}}
\hfill
\subfloat[Phase difference between $\mathscr{P}$ and $\mathscr{R}$]{\includegraphics[width=0.475\linewidth]{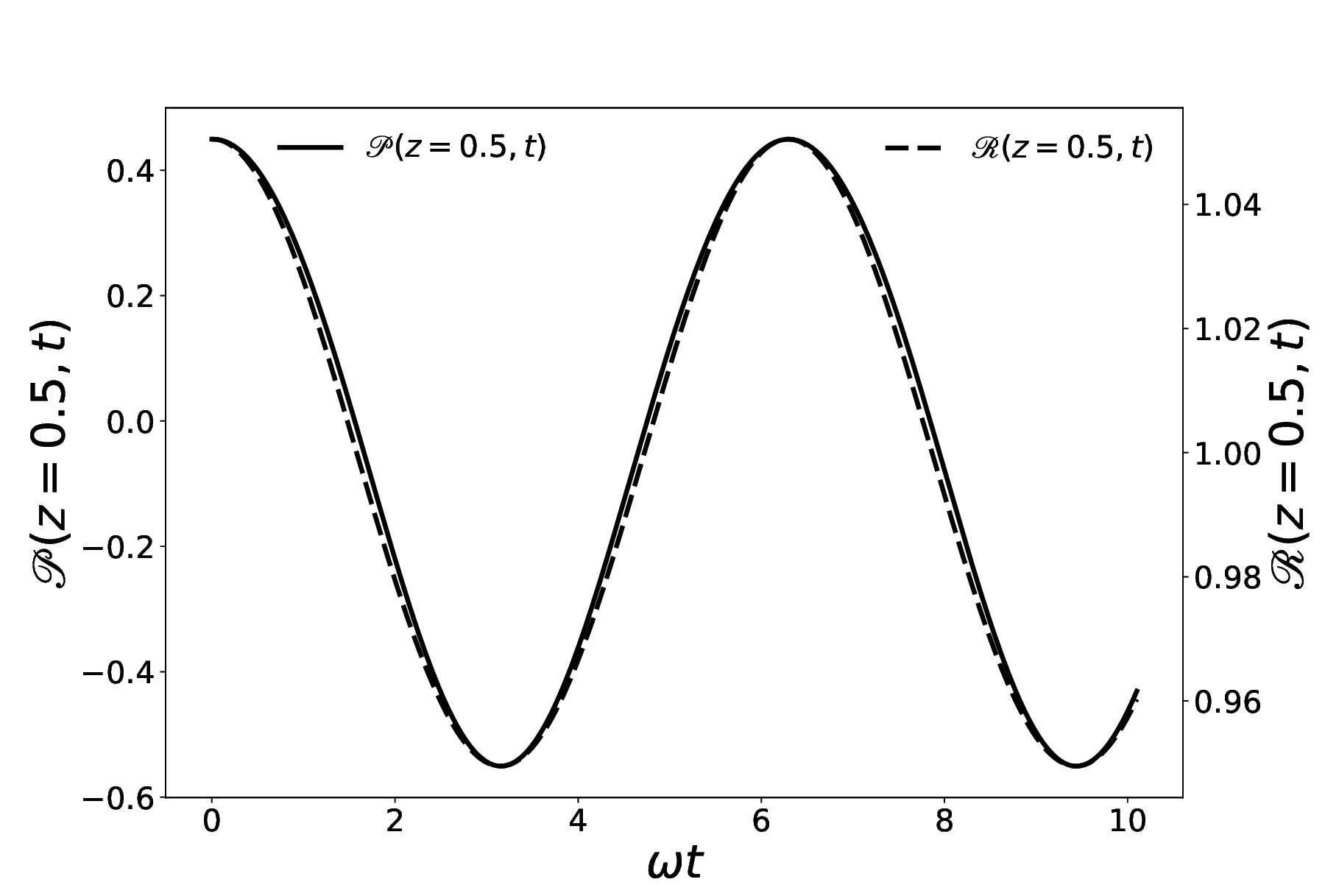}}
\caption{Standing waves due to FSI in a viscoelastic tube with  oscillatory pressure imposed at the inlet. (a) $\mathscr{R}(z,t)$ from Eq.~\eqref{eq:2_R_steady_6}. (b) $\mathscr{R}(z,t)$ and $\mathscr{P}(z,t)$ from Eq.~\eqref{eq:2_p_steady_3} at $ z = 0.5$, showing the phase difference $\phi$ between the two waves. Both plots are for $\De=10 $, $\St= 1$, $\alpha = 0.1$, and $\beta = 0.1$. The maximum phase difference $\phi$ between the deformed radius and the pressure is $\approx 0.5$ degrees. }
\label{fig:2_Standing_waves}
\end{figure}

From Eq.~\eqref{eq:2_R_steady_6}, we  observe that the tube radius maintains a phase difference with respect to the pressure imposed at the inlet. Due to the axial variation of the pressure, the tube radius has a maximum at the inlet, $z =0$, and a minimum at the outlet, $z = 1$. Combined with the oscillatory forcing, the tube wall sustains a standing wave; the inlet $z =0$ is antinode, whilst the outlet $z =1$ is a node, as shown in  Fig.~\ref{fig:2_Standing_waves}(a).

Our FSI theory neglects bending (and the boundary layers required to enforce the clamped boundary conditions at $z=0,1$). Thus, unlike the more common case of vibrating strings in musical instruments, the standing waves on the tube are not generated by the reflection of waves at the clamped ends. The standing wave pattern is the direct consequence of FSI. Specifically, the leading-order hydrodynamic pressure is itself in the form of a standing wave, as evidenced by Eq.~\eqref{eq:2_p^{0,0}}. The pressure standing wave then induces a deformation standing wave on the tube, at the same frequency but with a small phase difference (see Fig.~\ref{fig:2_Standing_waves}(b)). 

\begin{figure}
    \centering
    \includegraphics[width=0.5\linewidth]{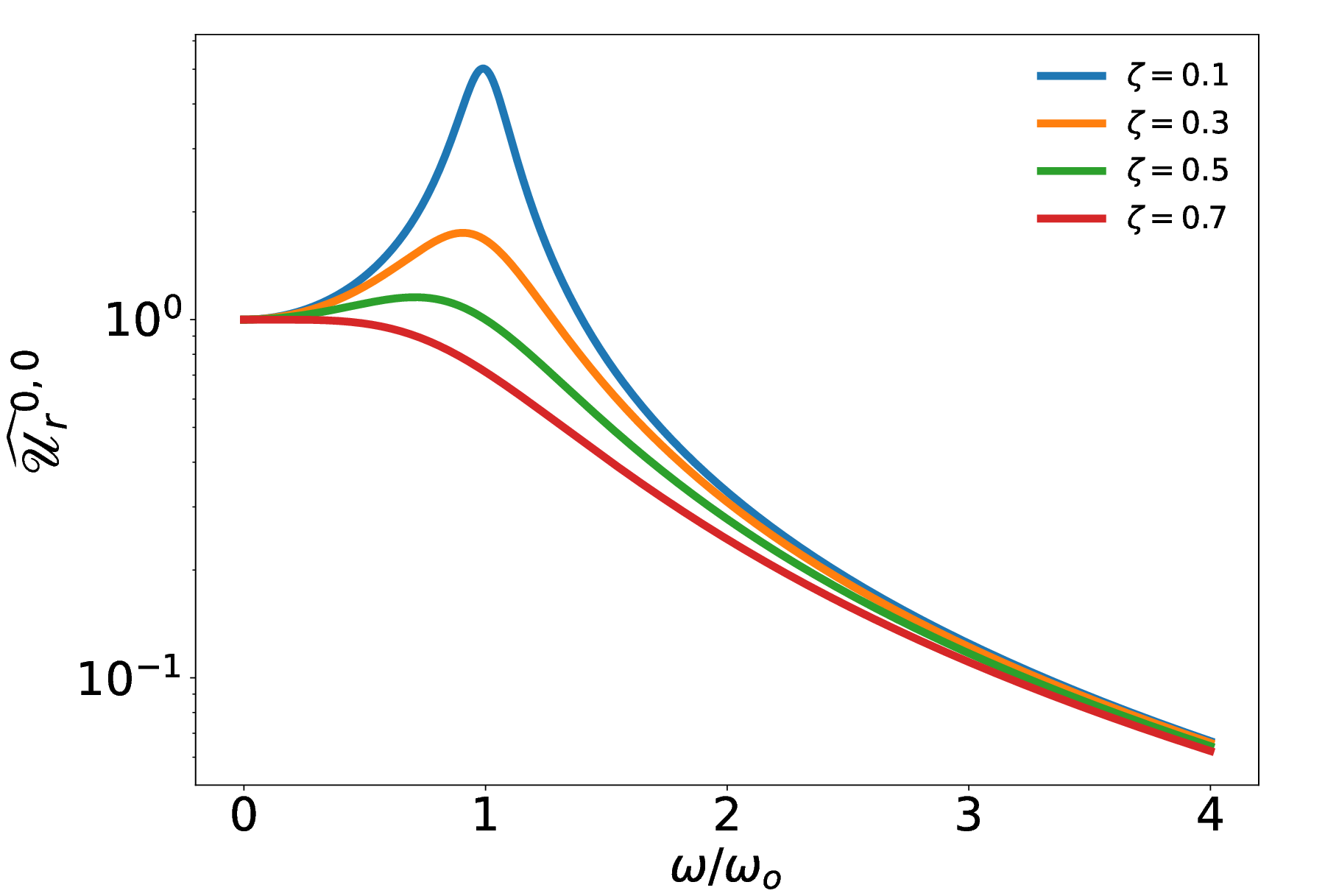}
    \caption{Frequency response of the post-transient radial deformation of the tube for different values of $\zeta = 1/(2\De\sqrt{\St})$, plotted from Eq.~\eqref{eq:R_amplitude}. Resonance occurs when the forcing frequency $\omega$ equals the resonant frequency $\omega_\mathrm{res}$ of the system, given by Eq.~\eqref{eq:2_Resonance_omega}. } 
\label{fig:Frequency_Response}
\end{figure}

To that end, it is possible, indeed desirable, to interpret Eq.~\eqref{eq:2_R_steady_6} in terms of the natural frequency of the system $\omega_{o} = 1/\sqrt{\St}$ and its critical damping ratio $\zeta = 1/(2\De\sqrt{\St})$:
\begin{subequations}\label{eq:R_phi_pt}
\begin{equation}
\label{eq:R_steady_2_omega}
    \mathscr{R}(z,t) = 1 +\beta \widehat{\mathscr{U}_r}^{0,0}(z,\omega)\sin{(\omega t +\phi)},
\end{equation}
where
\begin{equation}
\label{eq:R_amplitude}
    \widehat{\mathscr{U}_r}^{0,0}(z,\omega) = \frac{ (1-z)}{\sqrt{\left[1-\left({\omega}/{\omega_{o}}\right)^2\right]^2+4\zeta^2\left({\omega}/{\omega_{o}}\right)^2}}
\end{equation}
is the (positive) spatially varying amplitude of the tube deformation, and
\begin{equation}
    \phi = \sin^{-1}\left\{\frac{1-\left({\omega}/{\omega_{o}}\right)^2}{\sqrt{\left[1-\left({\omega}/{\omega_{o}}\right)^2\right]^2+4\zeta^2\left({\omega}/{\omega_{o}}\right)^2}}\right\}
\end{equation}
\end{subequations}
is the phase as before. Observe that $\widehat{\mathscr{U}_r}^{0,0}(z,\omega)$ reaches a maximum value at the ``resonant frequency''
\begin{equation}
    \label{eq:2_Resonance_omega}
    \omega = \omega_{\text{res}} =\omega_{o}\sqrt{1-2\zeta^2}.
\end{equation}
Therefore, the viscoelastic tube conveying oscillatory flow may be construed as a \emph{band-pass filter}, which allows signals close to $\omega_{\text{res}}$ to pass through, but attenuates signals with frequencies away from it, as shown in in Fig.~\ref{fig:Frequency_Response}. 

\subsubsection{Pressure response}
Next, we examine the fluid mechanical aspect of the problem. Similar to the case for impulsively pressurized inlet (Sec.~\ref{sec:result_impulsive}), the pressure profile for imposed oscillatory inlet pressure  exhibits exponentially decaying transients of $\mathcal{O}(\beta)$. Using the results from Sec.~\ref{sec:oscillatory}, we find that, once the transients die out, the post-transient pressure profile is
\begin{multline}
   \label{eq:2_p_steady_3}
   \mathscr{P}(z,t) = \left[(1-z) + 16\beta F(z)\hat{B}(\omega)\right]\cos{(\omega t)} - \left[8\alpha + 16\beta\hat{A}(\omega)\right]\omega F(z)\sin(\omega t) \\
  - \left[\beta {\hat{A}(\omega)}+\frac{\alpha}{4}\right]G(z)[1+\cos(2\omega t)]   -\beta\frac{\hat{B}(\omega)}{\omega}G(z)\sin(2\omega t).
\end{multline}
Note that this expression, at $\mathcal{O}(\alpha,\beta)$, has  higher harmonics of frequency $2\omega$, as well as time-independent terms due to mode couplings caused by FSI and compressibility.

Separating the different harmonics in  Eq.~\eqref{eq:2_p_steady_3}, we define the first harmonic:
\begin{multline}
 \label{eq:2_p_steady_harmonic_1_1}
 \mathscr{P}_{1H}(z,t) = (1-z)\cos{(\omega t)}-8\alpha\omega F(z)\sin{(\omega t)}\\
 + \beta\frac{16 \omega F(z) }{\sqrt{\left[1-\left({\omega}/{\omega_{o}}\right)^2\right]^2+{4\zeta^2}\left({\omega}/{\omega_{o}}\right)^2 }} \cos{(\omega t +\phi)},
\end{multline}
where $\phi$ is given in Eq.~\eqref{eq:2_R_steady_6}.
Importantly, the term of $\mathcal{O}(\beta)$ has a frequency-dependent amplitude with resonant frequency $\omega = \omega_{o}$ (the natural frequency of the system). Similarly, the second harmonic can be defined as
\begin{equation}
 \label{eq:2_p_steady_harmonic_2_2}
  \mathscr{P}_{2H}(z,t) =
  -\frac{\alpha}{4}G(z)\cos{(2\omega t)} - \beta \frac{G(z)  }{\sqrt{\left[1-\left({\omega}/{\omega_{o}}\right)^2\right]^2+{4\zeta^2}\left({\omega}/{\omega_{o}}\right)^2 }}\sin{(2\omega t +\phi)}.
\end{equation}
where $\phi$ is given in Eq.~\eqref{eq:2_R_steady_6}.  
Note that the envelope of the $\mathcal{O}(\beta)$ term in Eq.~\eqref{eq:2_p_steady_harmonic_2_2} has the same frequency dependence as that of radial deformation  (recall Eq.~\eqref{eq:R_steady_2_omega}) and, therefore, has the same resonant frequency $\omega_{\text{res}}$ given in Eq.~\eqref{eq:2_Resonance_omega}.

It is also instructive to note that, while both $\mathscr{P}_{1H}$ and $\mathscr{P}_{2H}$ exhibit similar frequency response, the spatial variations of their envelopes are  different. For $\mathscr{P}_{1H}$, the spatial variation of the envelope is set by $F(z)$, while for $\mathscr{P}_{2H}$, the spatial variation of the envelope is set by $G(z)$. From Eq.~\eqref{eq:Auxilliary}, since both $F(z)$ and $G(z)$ have the same zeros in $(0,1)$, the nodes of the two harmonics are also the same. On the other hand, the antinodes are different. They are at $z = (3-\sqrt3)/3$ for the first harmonic, and at $z =1/2$ for the second harmonic.

\subsubsection{Acoustic streaming}

Now, we discus the velocity field inside the tube, which is found from Eq.~\eqref{eq:velocity_axial_dimless} to be
\begin{equation}
\label{eq:velocity_axial_dimless_streaming}
 v_{z}(r,z,t) = -\frac{1}{4}\frac{\partial p}{\partial z}\left[ (1-r^2)+2\beta u_r\right] +\mathcal{O}(\beta^2).
\end{equation}
When we substitute the post-transient pressure profile from Eq.~\eqref{eq:2_p_steady_3} into Eq.~\eqref{eq:velocity_axial_dimless_streaming} , we obtain the post-transient axial velocity profile:
\begin{multline}
\label{eq:2_Post_Transient_V}
  \mathscr{V}_{z}(r,z,t) =\frac{(1-r^2)}{4}\Bigg\{  \left[1 - 16\beta F'(z) \hat{B}(\omega)\right] \cos(\omega t)
  + 8\left[\alpha F'(z)+2\beta\hat{A}(\omega)F'(z)\right]\omega\sin(\omega t) \\
  + \left[1+\cos(2\omega t)\right]\left[\frac{\alpha}{4} + \beta \hat{A}(\omega)\right]{G'(z)} 
  + \beta\frac{\hat{B}(\omega)}{\omega}G'(z)\sin{(2\omega t)}\Bigg\}\\
  + \frac{\beta}{4}\frac{(1-z)}{\sqrt{\left[1-\left({\omega}/{\omega_{o}}\right)^2\right]^2+{4\zeta^2}\left({\omega}/{\omega_{o}}\right)^2 }}\left[\sin(2\omega t+\phi)+\sin\phi\right].
\end{multline}

Observe that the velocity profile in Eq.~\eqref{eq:2_Post_Transient_V} consists of higher harmonics as well as a time-independent (steady-state) response. The time-independent part will lead to a nonzero mean flow (after averaging over a period of forcing). This observation leads us to predict \emph{acoustic streaming}, which is a common feature of oscillatory flows \citep{R01}. In the context of flows in tubes, a secondary nonzero mean flow has been observed for tubes with slowly varying radius and in curved tubes \citep{P80}. Such flows can also be generated by preset periodic boundary motion (peristalsis) for MEMS applications \citep{SS01}. On the other hand, for the problem considered herein, streaming is induced by FSI (under an oscillatory inlet pressure boundary condition).

\begin{figure}[t]
\centering
{\includegraphics[width=0.9\linewidth]{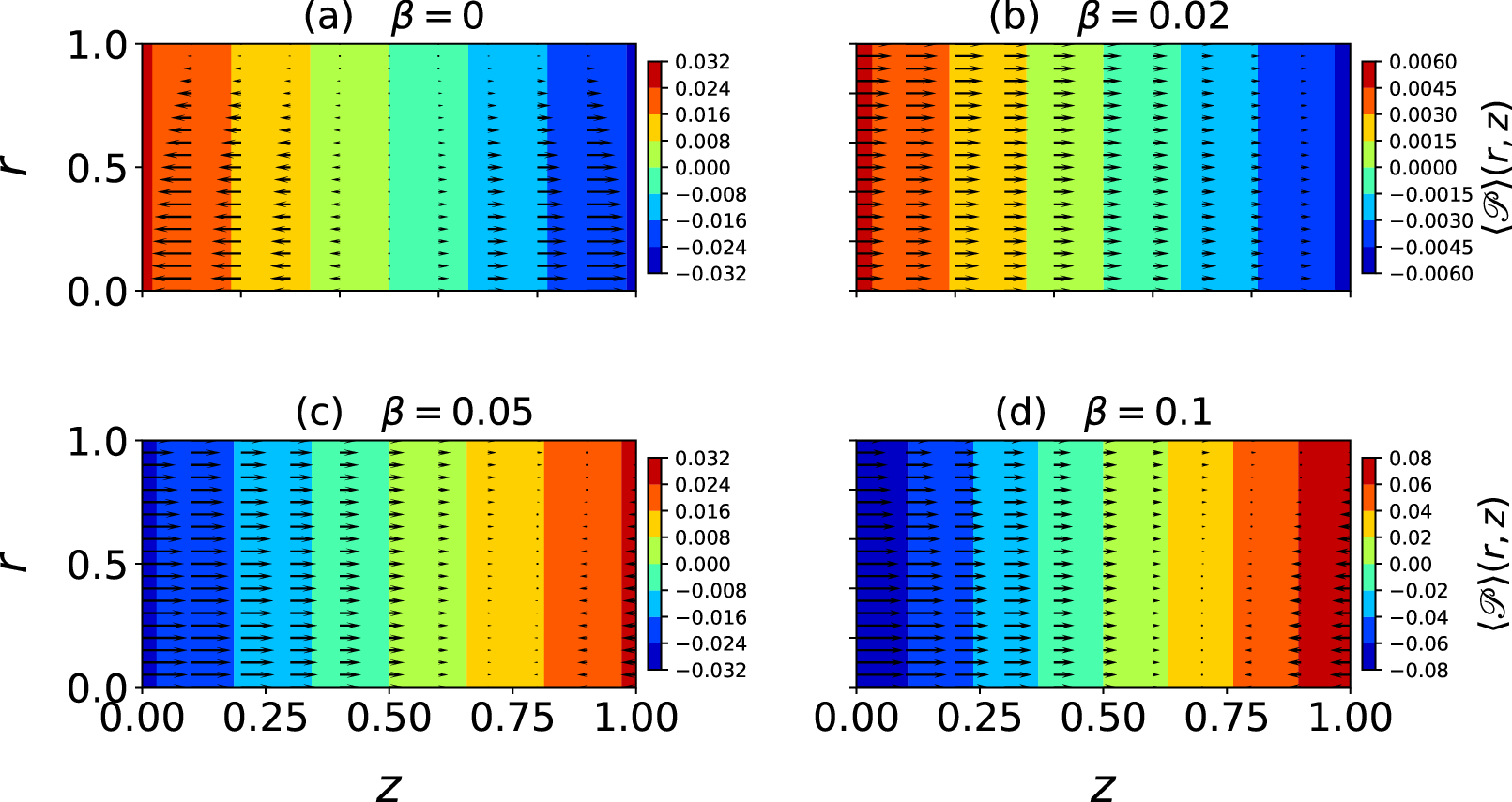}}
\caption{Quiver plots of the steady state acoustic streaming velocity $\langle \mathscr{V}_z \rangle(r,z)$ from Eq.~\eqref{eq:acoustic_velocity_streaming} superimposed onto contours of the quasi-steady part of the pressure $\langle \mathscr{P} \rangle(z)$ from Eq.~\eqref{eq:acoustic_radiation_2}, for different values of $\beta$ with $\alpha=0.1$ and $\omega/\omega_{o} =0.1$.} 
\label{fig:Acoustic_Streaming_Velocity_Profile}
\end{figure}

In an ideal fluid undergoing oscillatory flow, a fluid particle oscillates  about its position with a constant amplitude. Therefore, the average displacement of the fluid particle over a time period is zero. However, if a dissipative mechanism is introduced in the flow such that the restoring force acting on the fluid particle does not remain the same on both the sides of the mean position, then the mean position of the fluid particle undergoes a net (time averaged) displacement. Traditionally, the dissipative effects in the fluid have been introduced through viscosity (\textit{e.g.}, Rayleigh streaming \citep{Ray84}) or by a non-conservative body force \citep{R01}. However, our results show that it is also possible to generate a streaming flow through FSI. FSI couples the pressure gradient in the flow to the viscoelastic response of the tube, due to which the restoring force acting on the fluid particle varies, and a net displacement of the fluid particle occurs over a time period. 

To this end, the cycle-averaged acoustic streaming velocity is calculated to be
\begin{multline}
\label{eq:acoustic_velocity_streaming}
  \langle \mathscr{V}_z \rangle(r,z) = \frac{(1-r^2)}{4}\left\{\frac{\alpha}{4}+\beta \frac{1-\left({\omega}/{\omega_{o}}\right)^2}{\left[1-\left({\omega}/{\omega_{o}}\right)^2\right]^2+4\zeta^2\left({\omega}/{\omega_{o}}\right)^2}\right\}G'(z) \\ +\frac{\beta}{4}\frac{1-\left({\omega}/{\omega_{o}}\right)^2}{\left[1-\left({\omega}/{\omega_{o}}\right)^2\right]^2+4\zeta^2\left({\omega}/{\omega_{o}}\right)^2}(1-z),
\end{multline}
where $\langle\, \cdot\, \rangle \equiv \frac{1}{T}\int_{t}^{t+T} (\,\cdot\,)\,\rd t$, and, as before, $T = 2\pi/\omega$ is the period of oscillation. In the absence of FSI, $\beta =0$, and $\langle \mathscr{V}_z \rangle$ vanishes wherever $G'(z) = 0$, which are the anti-nodes of the streaming pressure (see Eq.~\eqref{eq:acoustic_radiation_2}). Therefore, the pressure anti-nodes coincide with streaming velocity nodes, as is common in acoustic streaming \citep{S12}. 
But due to FSI,  the relative location of pressure and velocity nodes/antinodes in the current streaming process changes. This trend is more clearly shown in Fig.~\ref{fig:Acoustic_Streaming_Velocity_Profile}. We note that for no FSI ($\beta = 0$), the velocity nodes are along the middle of the tube ($z =0.5$). However, for $\beta\ne0$, the position of the nodes shifts toward the tube's inlet ($z=0$). 

Perhaps, the most striking feature of the streaming velocity profile due to FSI, is the nonzero apparent slip velocity at $r = 1$. As seen in Fig.~\ref{fig:Acoustic_Streaming_Velocity_Profile}, the velocity  vectors have finite values at $r = 1$, for $\beta \neq 0$. This observation is also corroborated by Eq.~\eqref{eq:acoustic_velocity_streaming}. At a cursory level, this observation appears to be counter-intuitive since we explicitly enforced a no-slip boundary condition at the wall. However, this condition is imposed at the deformed wall, \textit{i.e.}, at $r = R(z,t) = 1+\beta u_r(z,t)$. But, since $\langle u_r(z,t) \rangle =0 $ and $\langle R(z,t) \rangle = 1$, the apparent boundary for the streaming velocity profile is at $r = 1$, where the no-slip has not been imposed and the streaming velocity is nonzero, appearing to slip.

\begin{figure}
    \centering
    \includegraphics[width=0.5\linewidth]{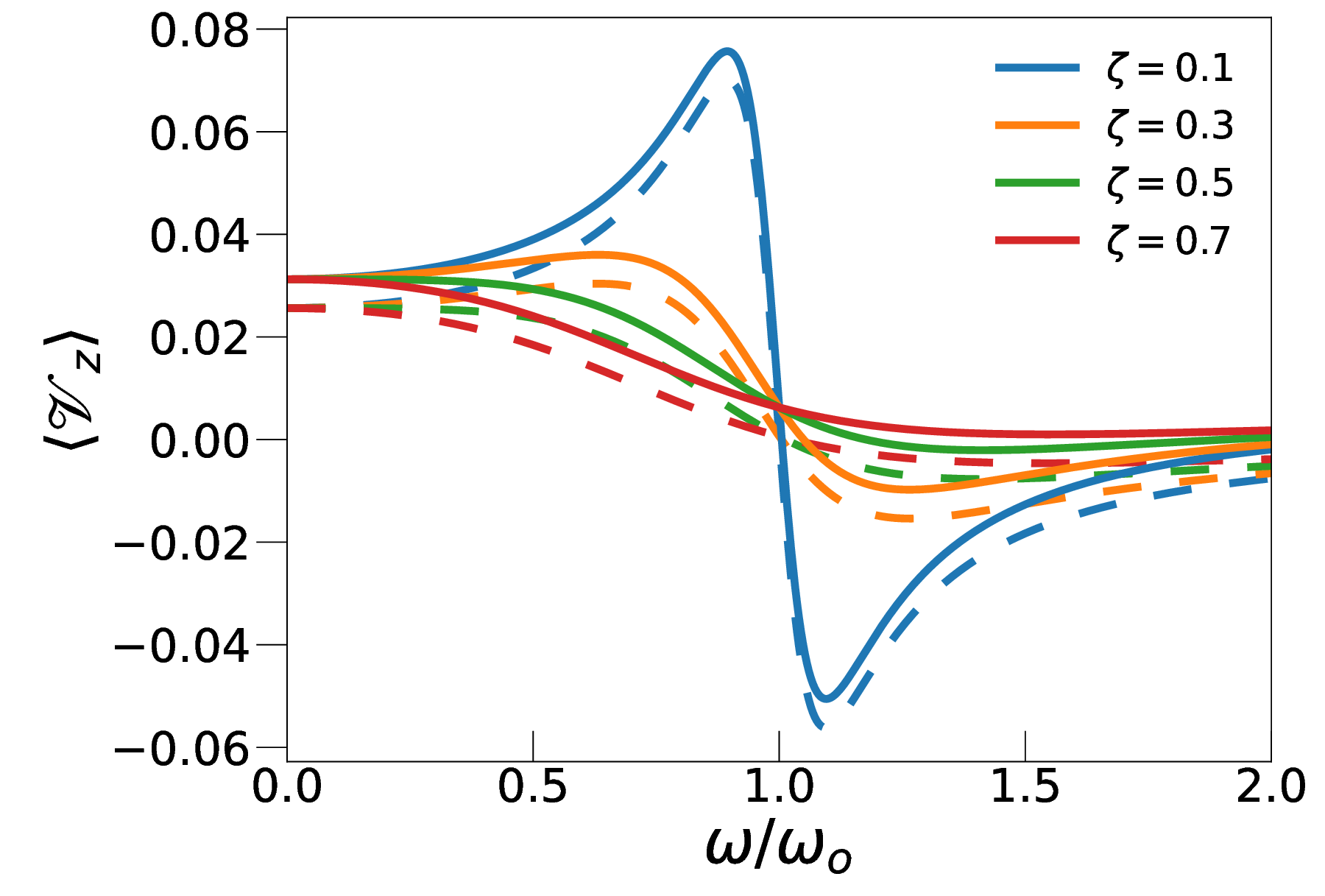}
    \caption{Frequency response of the period-averaged streaming velocity at $z =1$ and $r = 0$ from Eq.~\eqref{eq:acoustic_velocity_streaming}. The dashed curves denote $\alpha =0.01$, while the solid curves denote $\alpha = 0.1$.} 
\label{fig:Streaming_Velocity_Frequency_Response}
\end{figure}

It is also straightforward to deduce from Eq.~\eqref{eq:acoustic_velocity_streaming} that the (magnitude of) streaming velocity increases with increase with the compressibility number $\alpha$. On the other hand, an increase in the structural damping $\zeta= 1/(2\De\sqrt{\St})$, diminishes the (magnitude of) streaming velocity. This is shown more clearly in Fig.~\ref{fig:Streaming_Velocity_Frequency_Response}.

The stationary (time-independent) part of the pressure field is found from Eq.~\eqref{eq:2_p_steady_3} as:
\begin{equation}
 \label{eq:acoustic_radiation_2}
  \langle \mathscr{P} \rangle(z) = -\left[\frac{\alpha}{4}+\beta \hat{A}(\omega)\right]G(z)  =-\left\{\frac{\alpha}{4}+\beta \frac{1-\left({\omega}/{\omega_{o}}\right)^2}{\left[1-\left({\omega}/{\omega_{o}}\right)^2\right]^2+4\zeta^2\left({\omega}/{\omega_{o}}\right)^2}\right\}G(z).
\end{equation}
The nonzero mean pressure in an oscillatory field leads to an acoustic radiation force \citep{Lighthill2001}. This force can then be harnessed in many practical applications for, \textit{e.g.}, cell manipulation and droplet levitation, amongst other examples, which are studied under the umbrella of \emph{acoustophoresis} \citep{LL15}.
 
\begin{figure}
\centering
\subfloat[]{\includegraphics[width=0.49\linewidth]{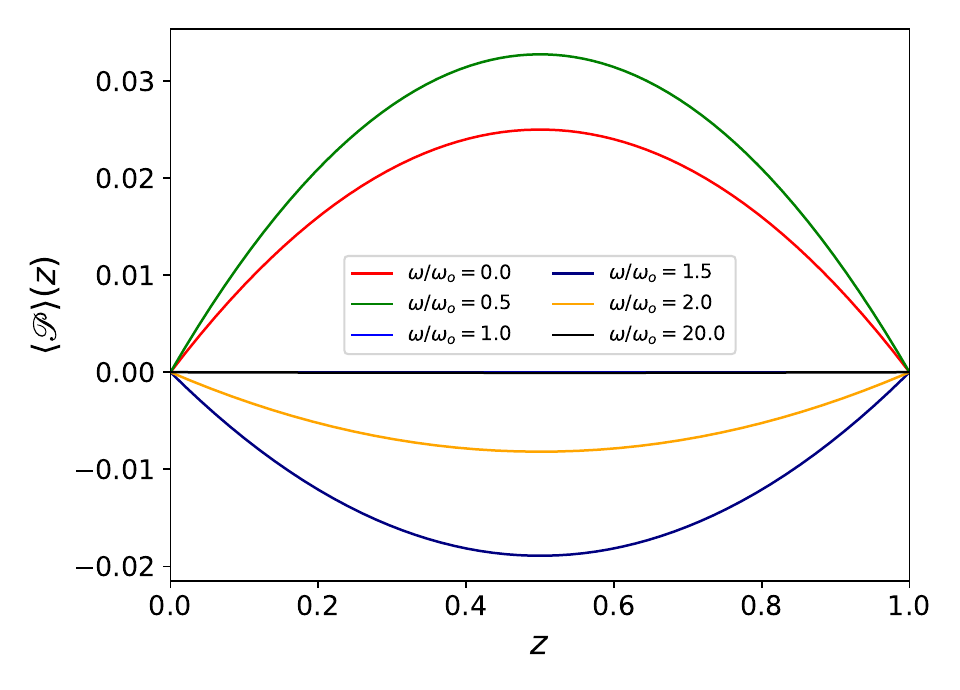}}
\hfill
\subfloat[]{\includegraphics[width=0.49\linewidth]{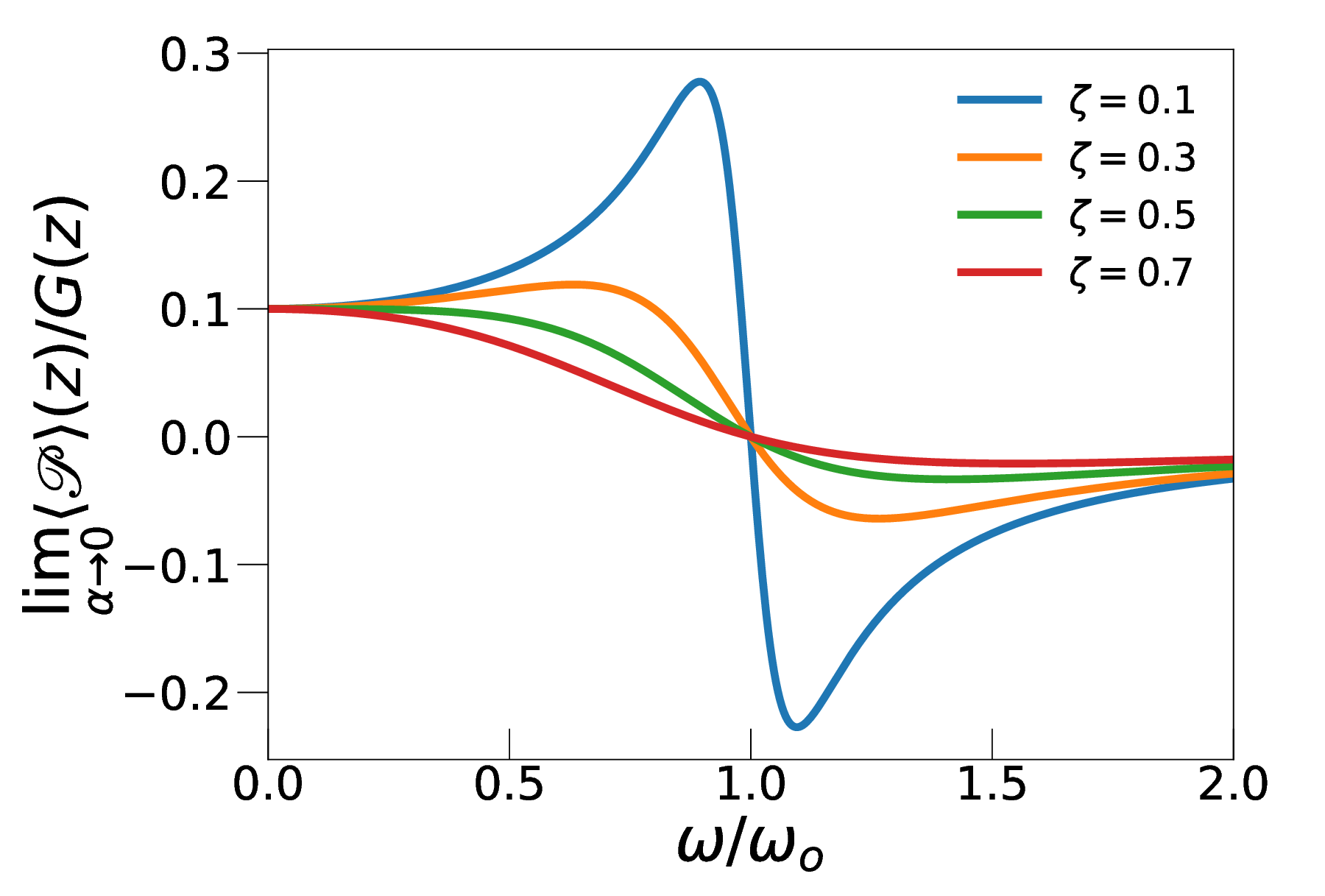}}
\caption{(a) The period-averaged acoustic radiation pressure profile $\langle \mathscr{P} \rangle(z)$ from Eq.~\eqref{eq:acoustic_radiation_2} for $\alpha = 0$ and $\beta =0.1$. (b) Frequency response of $\langle \mathscr{P} \rangle(z)$ for negligible compressibility $\alpha\to0$ and different values of $\zeta$, where $ \lim_{\alpha\to 0} \langle \mathscr{P} \rangle(z)/G(z)$ is evaluated from Eq.~\eqref{eq:acoustic_radiation_3}.}
\label{fig:Acoustic_Radiation_Pressure_Profile}
\end{figure}

An acoustic radiation force is thus expected to arise from $\langle \mathscr{P} \rangle(z)$ in Eq.~\eqref{eq:acoustic_radiation_2}, which has  a spatially varying  envelope given by $G(z)$.  The points of maximum pressure (\textit{i.e.}, the anti-nodes) are at $z$ such that $G'(z) =0$, \textit{i.e.}, at $z = 1/2$, while the points corresponding to zero pressure (\textit{i.e.}, the nodes), are at $z = 0,1$. These facts are illustrated in  Fig.~\ref{fig:Acoustic_Radiation_Pressure_Profile}(a). We also see from Eq.~\eqref{eq:acoustic_radiation_2}, that the $\mathcal{O}(\beta)$ term in the expression for $\langle \mathscr{P} \rangle(z)$ has a frequency-dependent response:
\begin{equation}
\label{eq:acoustic_radiation_3}
    \lim_{\alpha\to 0}\frac{\langle \mathscr{P} \rangle(z)}{G(z)} =-\beta \frac{1-\left({\omega}/{\omega_{o}}\right)^2}{\left[1-\left({\omega}/{\omega_{o}}\right)^2\right]^2+4\zeta^2\left({\omega}/{\omega_{o}}\right)^2}.
\end{equation}
Equation~\eqref{eq:acoustic_radiation_3} is plotted in Fig.~\ref{fig:Acoustic_Radiation_Pressure_Profile}(b), showing a decreasing (increasingly negative) trend as $\omega \to \omega_{o}$. However, near $\omega =\omega_{o}$, the profile undergoes a sharp rise to zero and becomes positive, before tapering off back to zero for $\omega \gg \omega_o$. The maximum magnitude is reached at $\omega = \omega_{o}\sqrt{1-2\zeta}$.  Clearly, the time-averaged pressure in a viscoelastic tube can be conceptualized as a low-pass filter, which allows  only the (pressure) signals with frequencies $\omega < \omega_{o}\sqrt{1-2\zeta}$ to pass through, and signals with higher frequencies are attenuated. We also observe from Fig.~\ref{fig:Acoustic_Radiation_Pressure_Profile} that, similar to the case of streaming velocity, the magnitude of streaming pressure also decreases with increase in structural damping.

The period-averaged post-transient volumetric flow rate enhancement, neglecting the terms of $\mathcal{O}(\beta^2)$, is calculated (from Eqs.~\eqref{eq:flow_rate_no_slip}, \eqref{eq:2_R_steady_6} and \eqref{eq:2_p_steady_3}) to be
\begin{multline}
    \langle q^*\rangle = \frac{1}{8}\left\{\frac{\alpha}{4} + \beta \frac{\left[1-\left({\omega}/{\omega_{o}}\right)^2\right] }{\left[1-\left({\omega}/{\omega_{o}}\right)^2\right]^2+4\zeta^2\left({\omega}/{\omega_{o}}\right)^2}\right\} G'(z)\\ 
    + \frac{\beta}{4}\frac{ (1-z)}{\sqrt{\left[1-\left({\omega}/{\omega_{o}}\right)^2\right]^2+4\zeta^2\left({\omega}/{\omega_{o}}\right)^2}}\sin{\phi}.
  \label{eq:q*}
\end{multline}
At the outlet ($z =1$), the second term in Eq.~\eqref{eq:q*} vanishes. Thus, for negligible compressibility ($\alpha\to0$), the nonzero mean enhanced flow at the outlet  has the same frequency response as the streaming pressure from Eq.~\eqref{eq:acoustic_radiation_3}. Consequently, the streaming-enhanced volumetric flow rate at the outlet also exhibits the low-pass filter response, with a cut-off frequency of $\omega_{o}\sqrt{1-2\zeta}$.

\section{Conclusion}
\label{sec:conclusion}

We analyzed compressible viscous flow at low Reynolds number in a compliant viscoelastic tube. The assumption of a slender geometry allowed us to neglect the convective inertia of the flow, leading to the lubrication approximation. The compressibility of the fluid was captured via an equation of state relating the density to the pressure via a compressibility parameter. For the structural mechanics problem, we employed an extension of the classical (linearly elastic) Donnell shell theory to incorporate Kelvin--Voigt (KV) linear viscoelasticity. We neglected bending, away from the clamped edges of the long and slender shell, and obtained a deformation equation, in which the dimensionless Deborah number $\De$ quantifies the relative magnitude of structural elasticity and structural viscosity, while a Strouhal number $\St$ quantifies the strength of (unsteady) inertial effects in the tube.

The coupled set of governing equations formulated for the fluid--structure interaction (FSI) problem were then solved in the context of three illustrative problems. First, steady-state compressible flow in an elastic tube was analyzed. The remaining two problems concerned transient dynamics. The first of these transient problems involved the impulsive pressurization of the tube's inlet, while the second transient problem involved an oscillatory pressure at the tube's inlet. While the steady-state problem reduced to a nonlinear separable ordinary differential equation that was solved exactly (but implicitly), the transient problems involved nonlinear partial differential equations, which required a double perturbation expansion, assuming weak FSI and weak compressibility, to make an analytical solution attainable.

We showed that, at steady state, both FSI and compressibility alter the velocity field, compared to incompressible flow in a rigid tube. In capillary viscometers, the classical Hagen--Poiseuille law is employed for viscosity estimation of gases. Neglecting compressibility may lead to incorrect estimation of the viscosity of a fluid by such devices \citep{BSG90}. Similarly, in microscale rheometry, a conduit constituted of a soft material like PDMS may ``pollute'' the viscosity estimation \cite{GWV16}, thus necessitating the use of our FSI theory.

The interplay between the key physics---FSI, compressibility, viscoelasticity, and inlet conditions---was analyzed. First, we found that, due to FSI, the Stokes flow in the tube experiences a transient adjustment to changes induced by the motion of the viscoelastic tube's walls. The dimensionless time constant given by the product $2\De\, \St$, which set the duration of the exponentially-decaying transient response of the system, was independent of the inlet boundary condition (impulsive or oscillatory pressure). Additionally, in the case of an oscillatory pressure imposed at the tube's inlet, FSI was shown to lead to acoustic streaming in the flow. The acoustic streaming induces an enhancement in the volumetric flow rate, which when averaged over a period of oscillation, displays a frequency response like that of a low pass filter with a dimensionless cut-off frequency of $\sqrt{1/\St-1/(\De\,\St^{3/2})}$ set by the dissipation ($\De$) and inertia ($\St$) of the tube. In particular, it was shown that oscillatory flow in viscoelastic tubes leads to resonance, which maximizes the deformation of the tube, at a particular value of the inlet pressure waveform's frequency. This result could have applications to transport, pumping and mixing \cite{SJW69} of gases in microfluidics \cite{SS01}.

Finally, although we solved the transient problem of compressible FSI in a viscoelastic tube using perturbation expansions, it may also worthwhile to explore another approach this problem by reducing the four coupled equations for $p$, $\rho$, $u_r$, and $q$ to a single nonlinear PDE in $u_r$, as discussed in Appendix~\ref{app:single_pde}. Additionally, the compressible flow analyzed in this work neglects any rarefaction (and wall slip) effects. Rarefaction is commonly encountered in gas flows when the characteristic length scale of the flow is comparable to the molecular mean free path of the gas \cite{ASB97,EJG18}. In future work, it may be worthwhile to analyze FSIs involving compressible flows with rarefaction (Knudsen number) effects.

\subsection*{Note added (arXiv v4)}
This version corrects an error in the integrated continuity equation~\eqref{eq:continuity_equation_integrated}, which propagated a factor of two into the coefficients of the transient and oscillatory pressure corrections [Eqs.~\eqref{eq:Terms}, \eqref{eq:continuity_equation_integrated_r}, \eqref{eq:continuity_equation_integrated_r2}, \eqref{eq:Fluid_Flow_LeadingOrder}, \eqref{eq:fluid_mechanics_first_order}, \eqref{eq:fluid_mechanics_first_order_b_soln}, \eqref{eq:1_p^{0,1}}, \eqref{eq:pressure_1_0_differential}, \eqref{eq:p^{1,0}}, \eqref{eq:2_p^{0,1}}, \eqref{eq:2_p^{1,0}}, \eqref{eq:pressure_transient}, \eqref{eq:2_p_steady_3}, \eqref{eq:2_p_steady_harmonic_1_1}, and \eqref{eq:2_Post_Transient_V}, and Appendices~\ref{app:expansions} and~\ref{app:single_pde}], and independently corrects the right-hand-side constant of Eq.~\eqref{eq:pressure_steady_general}, the sign of Eqs.~\eqref{eq:acoustic_radiation_2} and \eqref{eq:acoustic_radiation_3}, and Figs.~\ref{fig:1_R_St_De}(b), \ref{fig:1_R_A_Beta}(b), \ref{fig:2_Standing_waves}(b), \ref{fig:Acoustic_Streaming_Velocity_Profile}, and~\ref{fig:Acoustic_Radiation_Pressure_Profile}. No qualitative physical conclusion is affected, except that the streaming direction does not reverse with increasing $\beta$ (cf.\ Fig.~\ref{fig:Acoustic_Streaming_Velocity_Profile}). Full details are given in the Erratum \citep{AC26erratum}, in which equations and figures are numbered as in the published version of this article.

\subsection*{Acknowledgements}
This research was supported by the US National Science Foundation under grant No.\ CBET-1705637. We would also like to acknowledge the incisive and constructive feedback from the reviewers, which improved the manuscript.

\subsection*{Data Availability Statement}
The data that support the findings of this study are available within the article. Python script files will be made available on request via the Purdue University Research Repository.

\appendix
\section{Governing equations with perturbation expansions}
\label{app:expansions}

Substitution of the perturbation expansion from Eq.~\eqref{eq:perturbation_a} into Eq.~\eqref{eq:continuity_equation_integrated_r2} yields:
\begin{multline}
\label{eq:continuity_equation_integrated_r3}
  2\left(1+\alpha p^{ 0}\right) \left( R^{ 0}+\alpha R^{ 1}\right)\frac{\partial \left( R^{ 0}+\alpha R^{ 1}\right)}{\partial  t} +\frac{\partial  \left(1+\alpha p^{ 0}\right)}{\partial  t}\left( R^{ 0}+\alpha  R^{ 1}\right)^2 \\- \frac{\partial^2 \left(p^{ 0}+\alpha p^{ 1}\right)}{\partial z^2}\frac{ \left(1+\alpha p^{ 0}\right) \left( R^{ 0}+\alpha  R^{ 1}\right)^4}{8}-\frac{\left( R^{ 0}+\alpha  R^{ 1}\right)^4}{8}\frac{\partial  \left(1+\alpha p^{ 0}\right)}{\partial z}\frac{\partial \left(p^{ 0}+\alpha p^{ 1}\right)}{\partial z}\\-\frac{4\left( R^{ 0}+\alpha  R^{ 1}\right)^3}{8}\frac{\partial \left[ R^{ 0}+\alpha  R^{ 1}\right]}{\partial z}\frac{\partial \left(p^{ 0}+\alpha p^{ 1}\right)}{\partial z} \left(1+\alpha p^{ 0}\right)=0.
\end{multline}
At $\mathcal{O}(\alpha)$, Eq.~\eqref{eq:continuity_equation_integrated_r3} gives:
\begin{multline}
\label{eq:Fluid_Flow_FirstOrder}
    2\left( R^{ 1}\frac{\partial  R^{ 0}}{\partial  t}+\frac{\partial  R^{ 1}}{\partial  t} R^{ 0}+p^{ 0} R^{ 0}\frac{\partial  R^{ 0}}{\partial  t}\right) +
    \left( R^{ 0}\right)^2\frac{\partial p^{ 0}}{\partial  t} -\frac{\left( R^{ 0}\right)^4}{8}\left(4 \frac{ R^{ 1}}{ R^{ 0}}\frac{\partial ^2 p^{ 0}}{\partial z^2}+p^{ 0}\frac{\partial ^2 p^{ 0}}{\partial z^2}+\frac{\partial ^2 p^{ 1}}{\partial z^2}\right)\\
    -\frac{\left( R^{ 0}\right)^4}{8}\left(\frac{\partial p^{ 0}}{\partial z}\right)^2 
    -\frac{1}{8}\Bigg[12 \left( R^{ 0}\right)^2\frac{\partial p^{ 0}}{\partial z}  R^{ 1}\frac{\partial  R^{0}}{\partial z} + 4 \left( R^{ 0}\right)^3 \frac{\partial p^{ 0}}{\partial z}\frac{\partial  R^{ 1}}{\partial z}  \\
    + 4 \left( R^{ 0}\right)^3\frac{\partial p^{ 0}}{\partial z} \frac{\partial  R^{ 0}}{\partial z}p^{ 0}+4 \left( R^{ 0}\right)^3\frac{\partial p^{ 1}}{\partial z} \frac{\partial  R^{ 0}}{\partial z}\Bigg] =0.
\end{multline} 

Similarly, substituting the perturbation expansions from  Eqs.~\eqref{eq:perturbation_b2} into Eq.~\eqref{eq:Fluid_Flow_FirstOrder} yields:
\begin{multline}
\label{eq:Fluid_Flow_FirstOrder_b}
    2\left(\beta  R^{1 ,1}\right)\frac{\partial \left(1+\beta R^{0 ,1}\right)}{\partial  t}+2\frac{\partial \left(\beta  R^{1 ,1}\right)}{\partial  t}\left(1+\beta R^{0 ,1}\right)\displaybreak[3]\\
    +2\left(p^{0 ,0}+\beta p^{0 ,1}\right)\left(1+\beta R^{0 ,1}\right)\frac{\partial \left(1+\beta R^{0 ,1}\right)}{\partial  t} +
    \left(1+\beta R^{0 ,1}\right)^2\frac{\partial \left(p^{  0 ,0}+\beta p^{0 ,1}\right)}{\partial  t}\displaybreak[3]\\ -\frac{\left(1+\beta R^{0 ,1}\right)^4}{8}\Bigg[4 \frac{\left(\beta  R^{1 ,1}\right)}{\left(1+\beta R^{0 ,1}\right)}\frac{\partial ^2 \left(p^{0 ,0}+\beta p^{0 ,1}\right)}{\partial z^2}+\left(p^{0 ,0}+\beta p^{0 ,1}\right)\frac{\partial ^2 \left(p^{0 ,0}+\beta p^{0 ,1}\right)}{\partial z^2} \displaybreak[3]\\
    +\frac{\partial ^2 \left(p^{1 ,0}+\beta p^{1 ,1}\right)}{\partial z^2}\Bigg] 
    - \frac{\left(1+\beta R^{0 ,1}\right)^4}{8}\left[\frac{\partial \left(p^{0 ,0}+\beta p^{0 ,1}\right)}{\partial z}\right]^2 \displaybreak[3]\\
    -\frac{1}{8}\Bigg[12 \left(1+\beta R^{0 ,1}\right)^2\frac{\partial \left(p^{0 ,0}+\beta p^{0 ,1}\right)}{\partial z} \left(\beta  R^{1 ,1}\right)\frac{\partial \left(1+\beta R^{0 ,1}\right)}{\partial z} \\
    +4 \left(1+\beta R^{0 ,1}\right)^3 \frac{\partial \left(p^{0 ,0}+\beta p^{  0 ,1}\right)}{\partial z}\frac{\partial \left(\beta  R^{1 ,1}\right)}{\partial z}\\ +4 \left(1+\beta R^{  0 ,1}\right)^3\frac{\partial \left(p^{0 ,0}+\beta p^{  0 ,1}\right)}{\partial z} \frac{\partial \left(1+\beta R^{0 ,1}\right)}{\partial z}\left(p^{0 ,0}+\beta p^{0 ,1}\right)\\
    +4 \left(1+\beta R^{0 ,1}\right)^3\frac{\partial \left(p^{1 ,0}+\beta p^{1 ,1}\right)}{\partial z} \frac{\partial \left(1+\beta R^{0 ,1}\right)}{\partial z}\Bigg] =0.
\end{multline} 

\section{Reduction of the FSI problem to a single PDE}
\label{app:single_pde}

The FSI problem solved herein is completely characterized by four partial differential and algebraic equations namely Eqs.~\eqref{eq:flow_rate_defined},  \eqref{eq:continuity_equation_integrated},  \eqref{eq:constitutive_eq_dimless} and \eqref{eq:Deformation_ODE_final}. These equations among themselves govern the evolution of $q$, $p$, $u_r$ and $\rho$. In this paper, we have used double perturbation expansions to solve the coupled set of equations. On the other hand, it is also possible to reduce these four equations in four variables into a single equation in one variable, namely $u_r$. The nonlinear PDE can be shown to be:
\begin{multline}
\label{eq:single_equation_ur}
    \alpha\left(\frac{\partial u_{r}}{\partial t}+ \frac{1}{\De}\frac{\partial^2 u_{r}}{\partial t^2} + \St\frac{\partial ^3  u_{r}}{\partial  t^3}\right)\left(1+\beta u_r\right)^2 \\
    + \left[1+\alpha \left(u_{r}+ \frac{1}{\De}\frac{\partial u_{r}}{\partial t} + \St\frac{\partial ^2  u_{r}}{\partial  t^2}\right)\right]\left(1+\beta u_r\right)2\beta \frac{\partial u_r}{\partial t}\\ -\alpha\left(\frac{\partial u_{r}}{\partial z}+ \frac{1}{\De}\frac{\partial^2 u_{r}}{\partial t \partial z} + \St\frac{\partial ^3  u_{r}}{\partial  t^2 \partial z}\right)\left[\frac{1}{8}(1+\beta u_{  r})^4\right]\left[\frac{\partial u_{r}}{\partial z}+ \frac{1}{\De}\frac{\partial^2 u_{r}}{\partial t \partial z} + \St\frac{\partial^3  u_{r}}{\partial t^2 \partial z}\right] \\
    +\left[1+\alpha\left(u_{r}+ \frac{1}{\De}\frac{\partial u_{r}}{\partial t} + \St\frac{\partial^2  u_{r}}{\partial  t^2}\right)\right]\Bigg[-\left(\frac{\partial^2 u_{r}}{\partial z^2}+ \frac{1}{\De}\frac{\partial^3 u_{r}}{\partial t \partial z^2} + \St\frac{\partial ^4  u_{r}}{\partial  t^2 \partial z^2}\right)\frac{(1+\beta  u_{  r})^4}{8} \\
    -\left(\frac{\partial u_{r}}{\partial z}+ \frac{1}{\De}\frac{\partial^2 u_{r}}{\partial t \partial z} + \St\frac{\partial ^3  u_{r}}{\partial  t^2 \partial z}\right)\frac{(1+\beta  u_{  r})^3}{2}\beta\frac{\partial u_r}{\partial z}\Bigg] = 0,
\end{multline}
subject to the initial conditions
\begin{equation}
    u_r|_{t =0}  = 0 , \qquad \left.\frac{\partial u_r}{\partial t}\right|_{t = 0} = 0 ,
\end{equation}
and the boundary conditions
 \begin{subequations}
 \begin{align}
  \left.\left(u_{r}+ \frac{1}{\De}\frac{\partial u_{r}}{\partial t} + \St\frac{\partial ^2  u_{r}}{\partial  t^2}\right)\right|_{z =1} &= 0,  \\
  \left.\left(u_{r}+ \frac{1}{\De}\frac{\partial u_{r}}{\partial t} + \St\frac{\partial ^2  u_{r}}{\partial  t^2}\right)\right|_{z=0} &= H(t)\begin{cases}
  1, &\quad \text{(impulsive)},\\ \cos t, &\quad \text{(oscillatory)}.
  \end{cases}
  \end{align}
 \end{subequations}
 
Consistent with our perturbation expansion in the main text, collecting terms up to $\mathcal{O}(\alpha, \beta)$ and neglecting higher-order ones, we obtain the following simplified version of Eq.~\eqref{eq:single_equation_ur}:
\begin{multline}
    \alpha\left(\frac{\partial u_{r}}{\partial t}+ \frac{1}{\De}\frac{\partial^2 u_{r}}{\partial t^2} + \St\frac{\partial ^3  u_{r}}{\partial  t^3}\right) + 2\beta\frac{\partial u_r}{\partial t} -\frac{\alpha}{8}\left(\frac{\partial u_{r}}{\partial z}+ \frac{1}{\De}\frac{\partial^2 u_{r}}{\partial t \partial z} + \St\frac{\partial ^3  u_{r}}{\partial  t^2 \partial z}\right)^2 \\
    -\frac{\beta}{2}\frac{\partial u_r}{\partial z}\left(\frac{\partial u_{r}}{\partial z} + \frac{1}{\De}\frac{\partial^2 u_{r}}{\partial t \partial z} + \St\frac{\partial ^3  u_{r}}{\partial  t^2 \partial z}\right)\\ -\frac{1}{8}\left(\frac{\partial^2 u_{r}}{\partial z^2}+ \frac{1}{\De}\frac{\partial^3 u_{r}}{\partial t \partial z^2} + \St\frac{\partial ^4  u_{r}}{\partial  t^2 \partial z^2}\right)\left[1+4\beta u_r +\alpha\left(u_{r}+ \frac{1}{\De}\frac{\partial u_{r}}{\partial t} + \St\frac{\partial ^2  u_{r}}{\partial  t^2}\right)\right] = 0.
\end{multline}
However, this PDE is still nonlinear.
 
 
\bibliography{references_icc.bib}

\end{document}